\documentclass[]{article}

\usepackage[margin=1.3in]{geometry}

\usepackage{amsmath,amsthm,amssymb,euscript,mathtools} 
\usepackage{bm} 

\usepackage{graphicx,color,ae,fancyvrb,caption}
\usepackage[T1]{fontenc}

\usepackage{rotating,epstopdf}

\usepackage{setspace}
\usepackage{lineno}

\usepackage{booktabs, longtable, array}
\usepackage{enumitem}
\usepackage{array}

\usepackage{natbib}
\usepackage[colorlinks=true,allcolors=blue]{hyperref}

\let\cite\citep

\author{%
  Tobias Rüttenauer\textsuperscript{1,2,$\dagger$}, 
  Kasimir Dederichs\textsuperscript{2,$\dagger$}, 
  David Kretschmer\textsuperscript{2,$\dagger$}
  }

\date{\textsuperscript{1}Social Research Institute, University College London\\
\textsuperscript{2}Nuffield College, University of Oxford\\
\textsuperscript{$\dagger$} All authors contributed equally}

\title{Immigrant Residential Segregation in Europe: A Comparative Study of Spatial Segregation Patterns in Urban Areas across 30 Countries}

\begin{document}

\maketitle
{\renewcommand{\thefootnote}{\fnsymbol{footnote}}
\footnotetext[1]{Email: t.ruttenauer@ucl.ac.uk}
}

\begin{abstract}

Immigrant residential segregation can profoundly shape access to opportunities, immigrant integration, and inter-group relations. Yet we lack systematic evidence on how segregation varies across Europe, and what structural factors are associated with these patterns. This study addresses the gap by focusing on two questions: (i) how does immigrant-native segregation vary across urban areas in Europe, and (ii) which urban area- and country-level characteristics are consistently linked to segregation?

Using harmonised 1$\times$1 km grid-level data from the 2021/22 census, we calculate spatially weighted Dissimilarity Indices for all 717 Functional Urban Areas (FUAs) across 30 European countries. We combine these measures with rich data on demographics, the economy, housing, immigrant populations, and policy. To identify robust correlates of segregation, we apply a Specification Curve Analysis across 16,164 regression models.

Segregation is higher in Western and Northern Europe compared to most of Eastern and Southern Europe. Moreover, we show that segregation is heavily driven by macro-spatial dynamics between diverse urban cores and relatively homogeneous suburban areas. At the urban area level, segregation is systematically linked to the demographic composition and spatial distribution of the local population, economic conditions, housing market characteristics, as well as the composition of the immigrant population. At the national level, established immigrant destinations are more segregated, while migration and integration policies are not consistently linked to segregation.

These findings offer the most comprehensive comparative assessment of immigrant segregation across Europe to date, revealing how structural conditions relate to spatial integration.


\end{abstract}

\vfill

\begin{footnotesize}
Acknowledgements: The authors presented earlier versions of this paper at the annual/biennial meetings of the European Survey Research Association, the European Sociological Research Council, the Academy for Sociology, the German Sociological Association, the sociology and the migration group at Utrecht University, the migration group at the Berlin Social Science Centre, the Center for Critical Computational Studies at Goethe-University Frankfurt am Main, as well as seminars at Nuffield College and the UCL Social Research Institute. We would like to thank participants for their valuable feedback that improved the paper considerably.  
\end{footnotesize}

\thispagestyle{empty}
\clearpage
\onehalfspacing

\section{Introduction and Background}

Residential segregation has significant implications for the opportunities available to minority groups and for broader social cohesion. For instance, segregation restricts contact opportunities in neighbourhoods, limits the diversity of schools, community organizations, and public amenities \cite{Boterman.2016, Wiertz.2016}, and constrains access to resources and opportunities \cite{Chetty.2016, Sharkey.2014, Wodtke.2023}. A large body of research has documented racial segregation in the United States \cite{Elbers.2024, Lee.2008, Lichter.2015, Logan.1993, Logan.2025, Massey.1988, Roberto.2021} and immigrant segregation in European countries \cite{Benassi.2020, Catney.2023, Lichter.2019, Marcinczak.2023, Musterd.2005, Ruttenauer.2022, Sleutjes.2018, Spierenburg.2023, Teltemann.2015}. 

Traditionally, research has focused on three key causes of segregation: (i) preferences for living near in-group members \cite{Bruch.2006, Crowder.2008, Krysan.2009, Liebe.2023}, (ii) socio-economic disparities that lead to unequal housing budgets and sorting \cite{Alba.1999, Charles.2003, Lersch.2013}, as well as (iii) discrimination in the housing market \cite{Auspurg.2019, Desmond.2019, Faber.2020, Heath.2019, Quillian.2020, Rosen.2021}. While each cause is important in its own, recent work also highlights their interdependence and the role of structural barriers that simultaneously shape preferences, economic inequalities, and discriminatory practices \cite{Faber.2020, Krysan.2017, Lichter.2024, Parisi.2025}. This underscores the key role that structural urban characteristics may play in shaping segregation. Such urban area correlates of segregation have been the focus of several studies \cite{Benassi.2020, Iceland.2008, Kye.2023, Lichter.2024, Marcinczak.2023, Parisi.2025}. They have investigated how urban demographics, spatial layouts, institutional arrangements, and other characteristics are linked to segregation levels. In the US, studies document higher segregation in metropolitan areas with larger populations, higher minority shares, newer housing stock, and elevated poverty rates \cite{Iceland.2008, Kye.2023, Lichter.2024, Lichter.2015, Parisi.2025}. Comparative studies also underscore the role of the urban–suburban divide, with minorities often over-represented in urban cores \cite{Elbers.2024, Logan.2023}.

In Europe, comparative studies have similarly compared cities and explored structural correlates of immigrant segregation \cite{Andersson.2018, Benassi.2020, Benassi.2023, Johnston.2007a, Koopmans.2010, Lichter.2019, Marcinczak.2023, Musterd.2005}. These studies show, for example, that British cities tend to be more segregated than those in continental Europe, and that segregation is stronger among non-European vis-à-vis European immigrants \cite{Benassi.2020, Koopmans.2010, Marcinczak.2023, Musterd.2005}. Furthermore, findings suggest that segregation is stronger in urban areas with larger populations and a higher proportion of immigrants, and weaker in cities with wealthier and more educated populations \cite{Benassi.2020, Marcinczak.2023}.   

Although Europe offers a particularly compelling case for comparative research due to its diverse urban landscapes and diverse policy contexts, previous comparative studies are based on a set of only 5-8 European countries and data from 2011 or earlier. As a result, the majority of European countries – particularly those in Eastern and Southern Europe – have not yet been included in comparative studies, and recent immigration flows have not been taken into account. Moreover, each of the prior studies has considered only a narrow set of macro-level segregation correlates, preventing a comprehensive and up-to-date assessment of patterns and correlates of segregation across Europe. These limitations largely stem from data constraints, such as (a) inconsistent definitions of immigrant status across countries (e.g., citizenship vs. country of birth), (b) differences in the size and resolution of administrative spatial units, and (c) limited availability of harmonised country- and urban-level characteristics.

We address these limitations and provides a comprehensive analysis of immigrant residential segregation across Europe. We ask: (i) how does the level of immigrant segregation vary across European urban areas, and (ii) which characteristics of urban areas and national contexts are associated with segregation levels? Our analysis covers all 717 Functional Urban Areas (FUAs) across 30 European countries. FUAs are standardised spatial units that include city centres and surrounding commuting zones, comparable to metropolitan areas in the US. Using 1$\times$1 km grid cell data from the 2021/22 European census, we provide segregation indices for each FUA. The harmonised definition of spatial units and the harmonised measure of migration status ensure comparability across different national contexts in Europe. Our assessment of all urban areas across 30 countries significantly expands the geographical scope of existing research, for the first time mapping segregation patterns for most of the European continent. It also provides an important update to research based on the 2011 census.

We link these segregation indices to a rich, original dataset with information on 40 FUA- and country-level indicators drawn from Eurostat, the OECD, the UN, and other sources. This allows us to analyse how segregation is linked to the local population and its demographics, economic conditions, housing market characteristics, immigrant inflow and composition, and country-level policies and migration histories. 

Given the large number of covariates and plausible ways to specify models, we do not provide estimates for associations from a single and necessarily debatable model specification. Instead, we employ Specification Curve Analysis (SCA; Simonsohn, Simmons, and Nelson 2020) to estimate all 16,164  feasible and plausible regression models, and identify those structural factors which are consistently associated with immigrant segregation across these models. Our findings advance the understanding of how segregation is connected to local and national structural conditions and stimulates re-thinking existing theories about the emergence and consequences of segregation. We also provide a public dataset containing all segregation measures and macro-level indicators. This resource enables researchers, policymakers, and urban planners to explore and re-use our results for further research. 

\begin{sidewaysfigure}
    \centering
    \includegraphics[width=\textwidth]{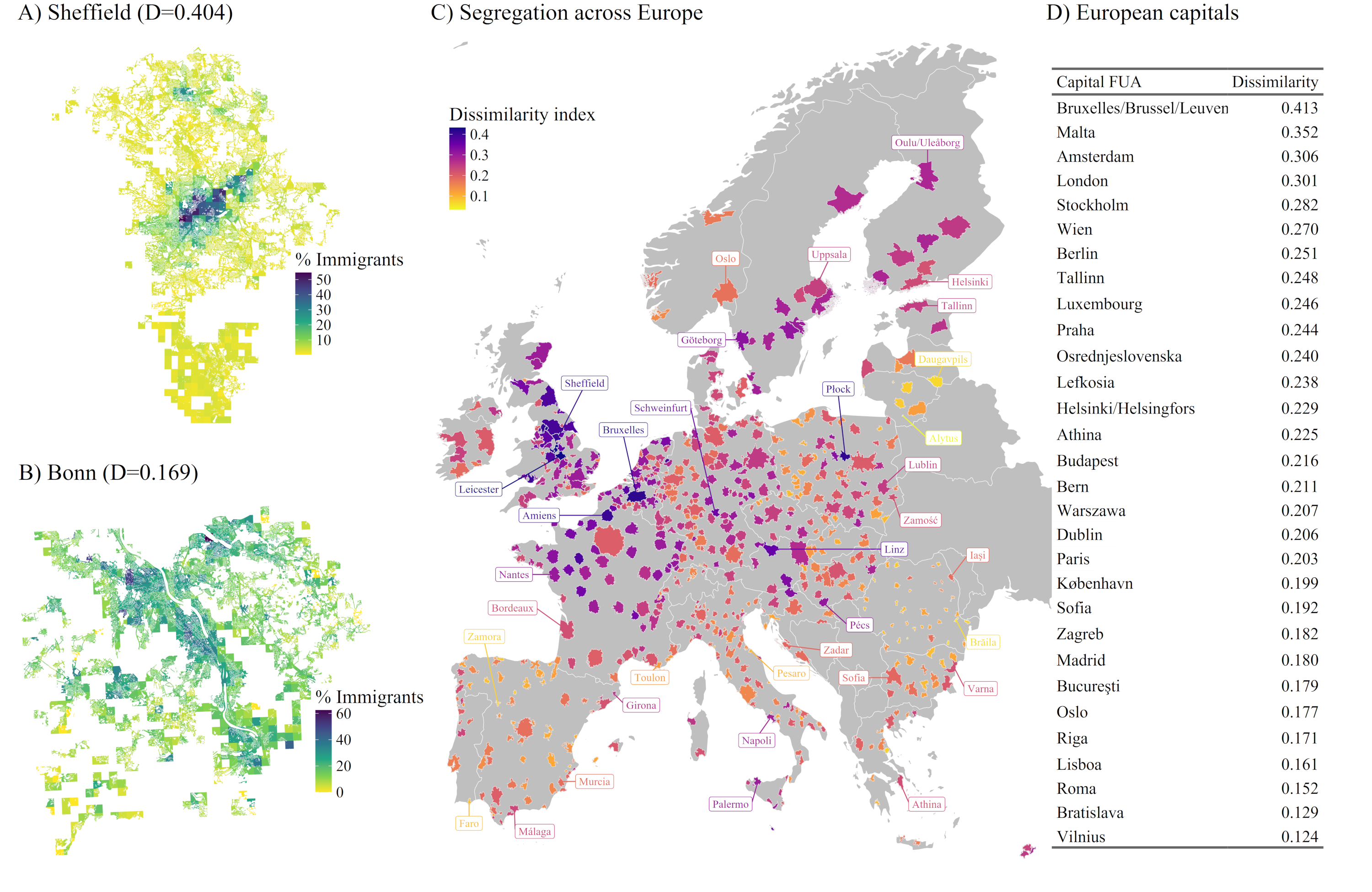}
    \caption{Spatial distributions of the immigrant population in A) Sheffield and B) Bonn; C) Residential segregation between natives and all immigrants in 717 European FUAs; D) Spatial dissimilarity index of each capital.}
    \label{fig:fig1}
\end{sidewaysfigure}

\section{Patterns of Segregation across Europe}

In our main results, we assess segregation with the spatial dissimilarity index $\widetilde{D}$ as proposed by \citet{Reardon.2008}. It measures `how different the composition of the individuals' local environments are, on average, from the composition of the population [in the FUA] as a whole' \cite[][see Methods section for more details]{Reardon.2008}; in other words, how unevenly immigrants and natives are distributed across space. We define the local environment with a 1km radius, which approximates institutional neighbourhoods within which most individuals can reach essential amenities \cite{Reardon.2008}.  We begin with the comparison of segregation levels (research question 1) and then analyse the correlates of segregation (research question 2). 

\textbf{Mapping Segregation Levels Across Europe}. To illustrate our measure of segregation, Figures \ref{fig:fig1} A) and B) depict the variation of the immigrant share per 1$\times$1 km grid cell for the two FUAs of Leicester (United Kingdom) and Bonn (Germany). In Leicester, the proportion of immigrants varies widely across grid cells. Those grid cells with high proportions of immigrants are spatially clustered around the urban core, while natives are overrepresented in the suburban surroundings, indicating substantial segregation. This is reflected in Leicester's dissimilarity index of 0.41 – one of the highest in Europe. In Bonn, by contrast, the proportion of immigrants varies less across grid cells, and grid cells with high proportions of immigrants are distributed across the entire FUA, including its suburban areas. Correspondingly, the spatial dissimilarity index for Bonn is 0.17, and thus significantly lower than in Leicester. See supplementary Figure A1 for additional examples.

\begin{figure}[h!t]
    \centering
    \includegraphics[width=\textwidth]{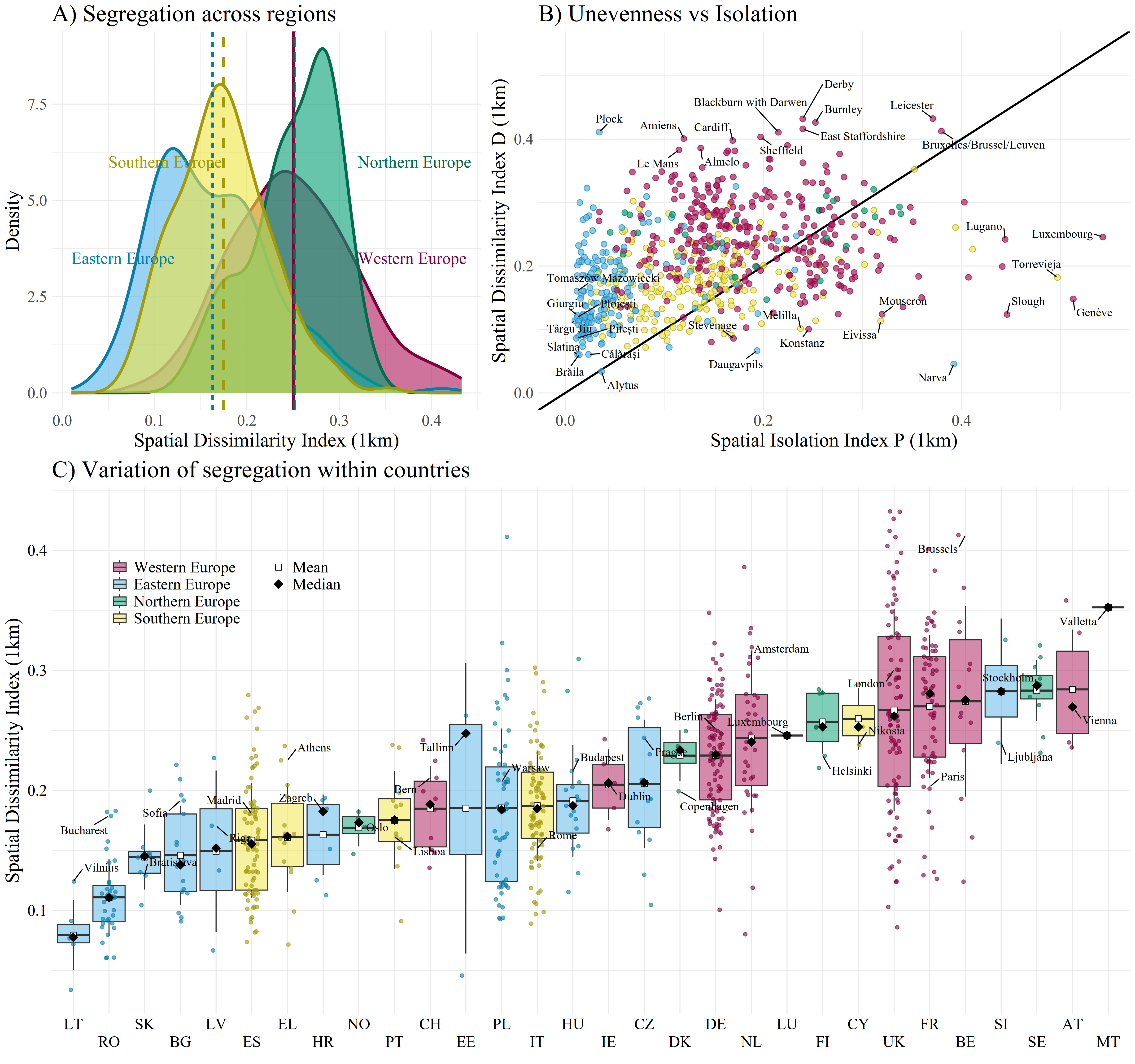}
    \caption{Segregation across 717 FUAs across 30 countries measured by the spatial Dissimilarity Index. A) Density distribution of segregation across regions; B) Comparison with spatial Isolation Index; C) Residential Segregation of FUAs and country boxplots with capitals highlighted. Note: The box of each boxplot represents the interquartile range (IQR) across FUAs and the whiskers extend to the smallest and largest values within 1.5xIQR. Countries are ordered by the mean of the segregation index across their FUAs. }
    \label{fig:fig2}
\end{figure}

Figure \ref{fig:fig1} C) maps the spatial dissimilarity index $\widetilde{D}$ across all 717 FUAs, revealing that segregation is stronger in Western and Northern vis-à-vis Eastern and Southern Europe (see also Figure \ref{fig:fig2} A). The FUAs with the highest levels of residential segregation are primarily located in the United Kingdom, northern France, and Belgium. Segregation is also substantial in the Netherlands, Germany (particularly East Germany), Denmark, Poland, Sweden, and Finland. FUAs in most Eastern and Southern European countries are considerably less segregated. There are, however, two notable exceptions: Norwegian FUAs are less segregated than those in other Northern European countries, and Polish FUAs are more segregated than those in other Eastern European countries. 

\textbf{Comparing Patterns of Segregation across Europe}. Figure \ref{fig:fig2} C) summarizes the distribution of these segregation indices in a boxplot for each country, with dots representing individual FUAs and capitals highlighted in black. Countries are ordered by their median segregation level, mirroring the Western/Northern vs. Eastern/Southern gradient in segregation (see also Figure \ref{fig:fig1} C and \ref{fig:fig2} A). Roughly 40\% of the total variation in segregation occurs between and 60\% within countries, suggesting that both country- and urban-level characteristics are relevant for segregation. Within-country variation is particularly visible in the five most populous countries (Germany, France, United Kingdom, Italy, Spain), where the gap between the most and the least segregated FUA exceeds 0.2 scale points, i.e., half of its entire range. The variation is strongest in England, where the dissimilarity index ranges from 0.09 in Stevenage to 0.43 in Leicester. Segregation ranges from 0.13 (Cannes-Antibes) to 0.40 (Amiens) in France, and from 0.10 (Konstanz) to 0.35 (Schweinfurt) in Germany. Variation is smaller in Spain and Italy, where we measure segregation levels from 0.07 (Zamora) to 0.27 (Las Palmas) and 0.10 (Pesaro) to 0.30 (Palermo), respectively. 

Most capitals (Figure \ref{fig:fig1} D) show higher segregation levels than their respective national average. Large immigrant destinations in Western and Northern Europe, like Brussels (0.41), Amsterdam (0.31), London (0.30), and Stockholm (0.28), have higher segregation levels than most other capitals. The least segregated capitals at the 1km scale can be found in Eastern and Sourthern Europe: Vilnius (Lithuania, 0.12), Bratislava (Slovakia, 0.13), and Rome (Italy, 0.15). Madrid (Spain, 0.18), Paris (France, 0.20), Berlin (Germany, 0.25), and other capitals fall in between. The full set of capitals and their segregation levels are shown in Figure \ref{fig:fig1} C. 

Across most urban areas, non-EU immigrants live more segregated than EU immigrants (see Figure \ref{fig:degree1} A). Locations close to the EU border (such as Chelm or Zamość) are exceptional in this aspect, as non-EU immigrants might be relatively close to the native population in terms of language and cultural heritage in these places. Segregation indices of FUAs closer to the EU borders (like Larisa, Trikala, or Siedlce) are also more affected by outlier grid cells that likely contain facilities for asylum seekers. Nevertheless, the general impact of such outliers is very small (see Figure \ref{fig:degree1} B).

\textbf{Dimensions of Segregation}. As shown by previous research, segregation is a multidimensional phenomenon \cite{Lichter.2015, Massey.1988, Reardon.2008, Reardon.2004, Ruttenauer.2022, Spierenburg.2023}. For instance, besides unevenness (captured with the dissimilarity index $\widetilde{D}$), another dimension of segregation is immigrants' isolation from natives ($\widetilde{P}$). Figure \ref{fig:fig2} B plots these dimensions against each other (see also Figure \ref{fig:fig2_p}). For most FUAs, immigrant isolation is lower than dissimilarity, as indicated by their location above the 45 degree line. In particular, many Eastern European FUAs score high on spatial dissimilarity but low on isolation (e.g. Zamość or Płock). This reflects the relatively low number of immigrants living in these urban areas and the strong dependency of the isolation index on the immigrant population. In fact, the size of the immigrant population explains 96\% of the variation in $\widetilde{P}$. Net of immigrant population, however, the isolation index $\widetilde{P}$ is strongly correlated with $\widetilde{D}$ (r = 0.79). 

In line with previous research \cite{Lichter.2015, Reardon.2008}, segregation indices decline when measured at larger spatial scales (e.g., 2km or 5km see Figure \ref{fig:degree2}, panel A). Although the overall regional pattern is unaffected (see Figures \ref{fig:fig2_500m} and \ref{fig:fig2_5km}) and segregation levels generally are highly correlated across spatial scales (e.g., r = 0.87 for 1km and 5km), there is interesting geographical variation in the scale dependency of segregation levels. Specifically, several urban areas in Eastern and Southern European countries are much less segregated at larger spatial scales as compared to the 1km level (see Figure \ref{fig:scale_box}). Many urban areas in Western and Northern Europe, on the other hand, exhibit high levels of segregation even at broader spatial scales. 

\begin{figure}[ht!]
    \centering
    \includegraphics[width=\textwidth]{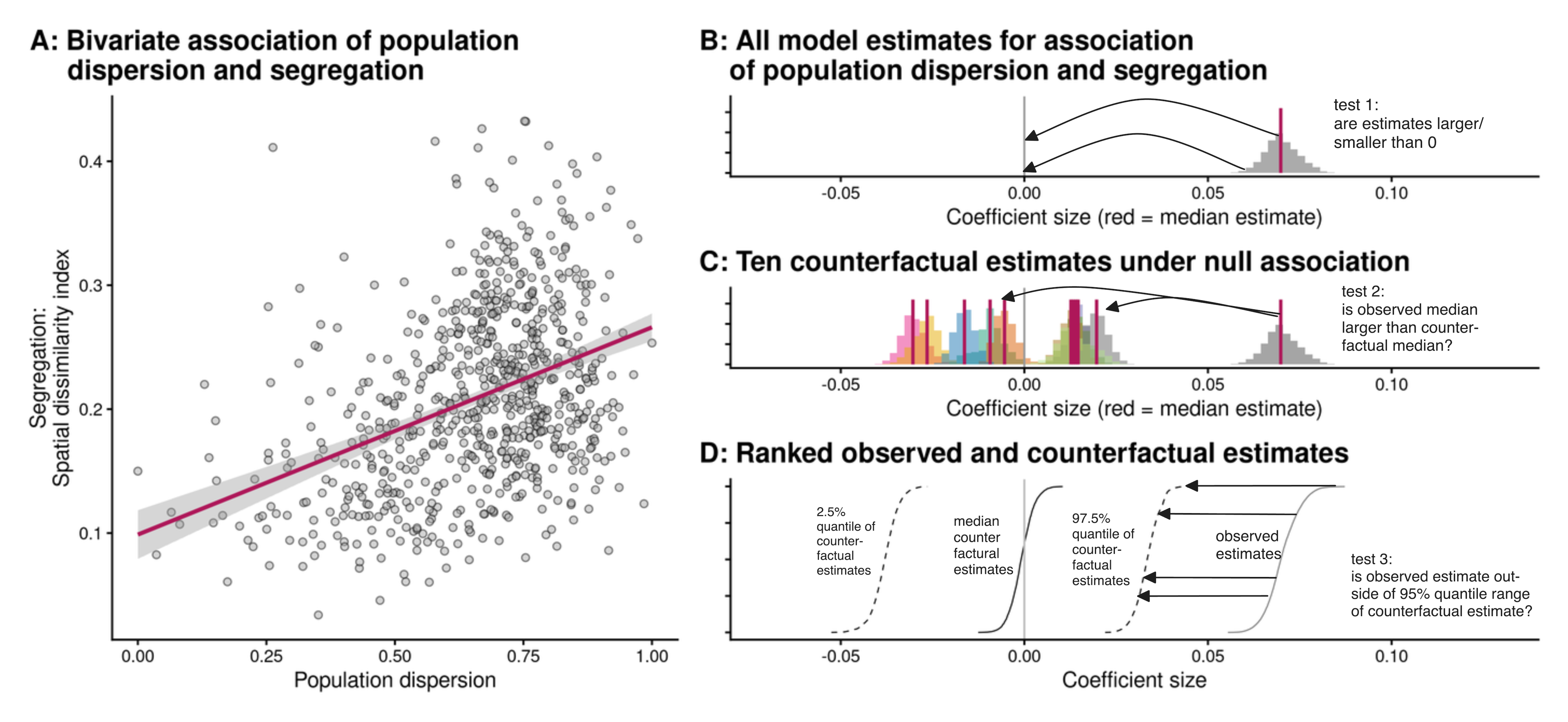}
    \caption{Example analysis of the association of levels of segregation with FUA population dispersion: (A) Bivariate association; (B) Histogram of coefficients from Specification Curve Analysis; (C) Counterfactual coefficient distribution under null association; (D) Ranked observed vs. counterfactual estimates. Note: In panel D, the dashed lines correspond to the 95\% quantile band of the ranked counterfactual bootstrapped estimates. The grey line corresponds to the original estimates.}
    \label{fig:fig3}
\end{figure}

Rather than the spatial scale itself, a key driver of segregation in Western European FUAs is the stark contrast in diversity between urban cores and their surrounding suburban zones. When comparing dissimilarity indices for entire FUAs with those calculated for core cities only (see Figure \ref{fig:degree2}, panel B), we find that most cities are less segregated within the urban core than across the whole commuting zone. Examples include some of the most strongly segregated capitals: for instance, the spatial dissimilarity index for the Brussels urban area is 0.41, while its core city only exhibits a dissimilarity index of 0.15. Likewise, Amsterdam has an overall dissimilarity of 0.31, while its core city only exhibits a value of 0.12. This urban–suburban divide closely mirrors findings from the US context \cite{Elbers.2024, Lichter.2015, Logan.2023, Logan.2025}, suggesting that large-scale segregation between inner cities and suburban commuting zones characterises North American and European metropolitan regions. 

\section{Urban- and Country-Level Correlates of Segregation}

We next examine which characteristics of urban areas are systematically linked to segregation using our core measure: the spatial dissimilarity index with a 1km radius. Since we consider a wide range of potentially interrelated characteristics, we do not estimate a single `most likely' omnibus model for these associations. Instead, we borrow from the logic of specification curve analysis (SCA), estimating all plausible and feasible model specifications to determine which associations are robust across these specifications \cite{Simonsohn.2020}. In total, we estimate 9,990 model specifications which all include five FUA population controls (population size, density, concentration, immigrant proportion, and educational composition), up to four varying additional FUA-level correlates, and country fixed effects. For each correlate of segregation, this yields a distribution of coefficients. Three increasingly conservative tests assess the robustness of associations across specifications (see Methods section for more information). 

Figure \ref{fig:fig3} illustrates this approach for the association between segregation and population dispersion, which captures how spatially dispersed the overall population is across the FUA. The bivariate association in panel A suggests that stronger population dispersion is linked to higher levels of segregation. Panel B displays the distribution of estimated associations across all specifications from the SCA, which differ in the set of covariates included next to population dispersion. Test 1 evaluates whether at least 95\% of all specifications predict the same sign for the variable's association with segregation. Tests 2-3 further compare the observed distribution of estimates with 500 counterfactual bootstrapped distributions that enforce a null association between the focal variable and segregation \cite{Simonsohn.2020}. Next to the observed coefficients in grey, panel C illustrates ten example counterfactual distributions. Test 2 assesses whether the median observed estimate is larger than the median in at least 95\% of the 500 bootstrapped distributions. Panel D visualizes the results of test 3, which considers each model specification separately and ranks the estimates across specifications by size. For each specification, we test if the observed estimate falls outside of the 95\% quantile band of ranked counterfactual bootstrapped estimates (dashed lines), indicating that the observed estimate will almost never be induced if there is no actual association. An association passes this test if observed estimates fall outside of the quantile band for at least 95\% of all specifications. As this third test is very conservative, we also label associations as robust that only pass tests 1 and 2 but not test 3.  

\begin{figure}[h!]
    \includegraphics[width=\textwidth]{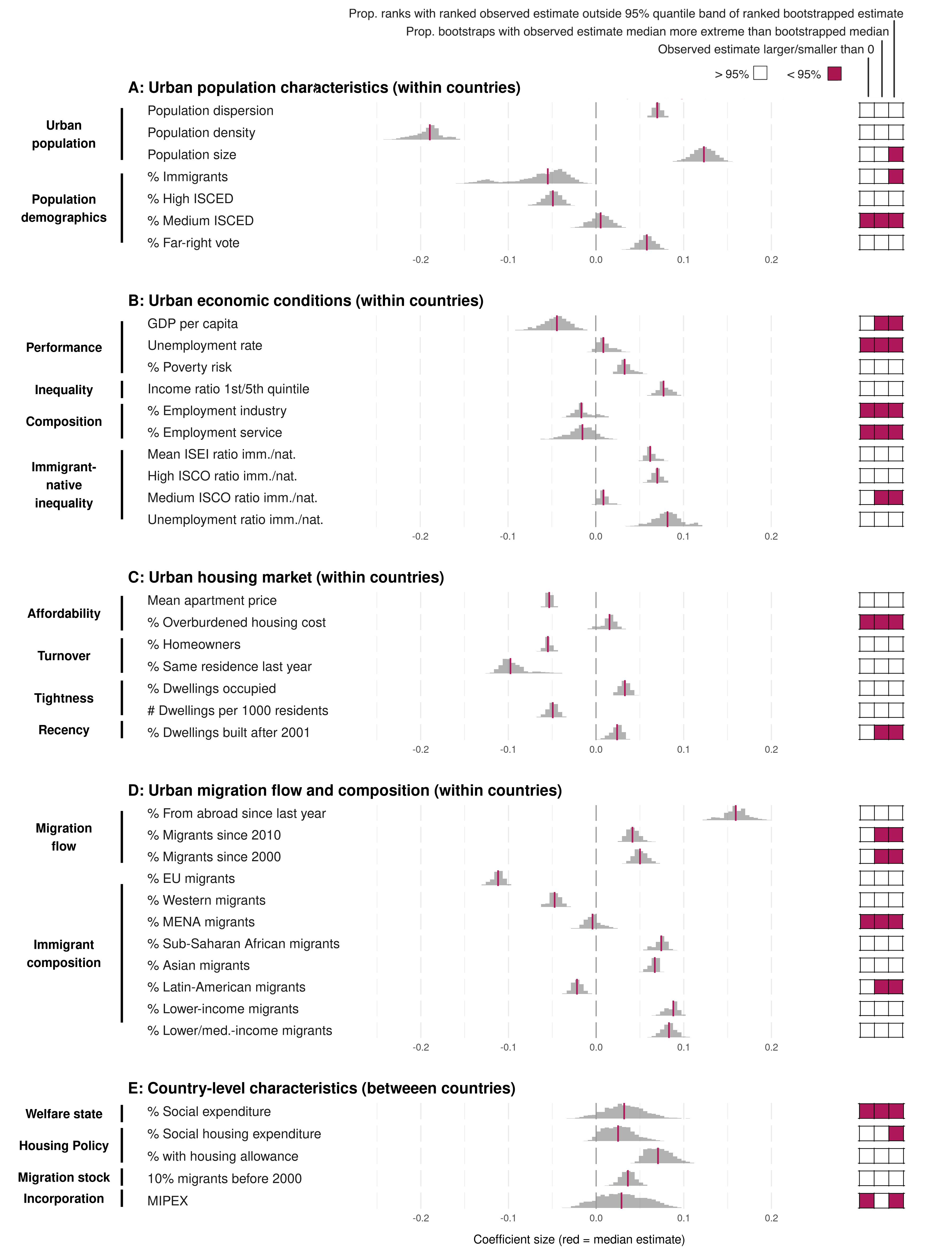}
    \caption{Specification Curve Analysis of associations of levels of segregation with (A) FUA population characteristics, (B) FUA economic conditions, (C) FUA housing market conditions, (D) FUA immigration flow and composition, (E) country-level characteristics. Note: The universe of `reasonable' model specifications excludes specifications including multiple covariates from the same covariate subgroup. For example, we do not estimate models that include both GDP per capita and unemployment rate, two indicators in the economic performance covariate subgroup.}
    \label{fig:fig4}
\end{figure}

The association between population dispersion and segregation in Figure \ref{fig:fig3} passes all three tests. This provides robust support for the idea that segregation of immigrants is stronger in FUAs whose population is distributed more evenly across the entire urban area. When most of the population lives very concentrated in only a few parts of the city, immigrant segregation is lower. Figure \ref{fig:fig4} extends this example to the complete set of covariates, grouped into five thematic domains: population characteristics (A), economic conditions (B), the housing market (C), migration flow and immigrant composition (D), and country-level characteristics (E). Except for the country-level characteristics, associations are estimated using models with country fixed effects, i.e., models that only consider variation within rather than across countries.

\textbf{Panel A: Population characteristics}. Next to higher population dispersion, also lower population density is robustly associated with higher levels of segregation across all statistical tests. Furthermore, FUAs with larger populations are more segregated in all specifications, though the robustness of this link does not extend to our most conservative statistical test. Regarding population demographics, a higher proportion of immigrants is associated with lower segregation, but effect sizes vary widely across specifications. All tests further support that a larger proportion of highly educated residents and a smaller group of right-wing voters correlate with lower levels of segregation.

\textbf{Panel B: Economic conditions}. Favourable economic conditions, i.e., higher GDP per capita, lower poverty risk, and lower unemployment rate, are all linked to lower segregation levels, but only poverty risk is robustly associated across all statistical tests. Immigrant segregation is also consistently higher in FUAs with greater economic inequality and in FUAs where immigrants’ unemployment rates exceed natives’ more strongly. Somewhat surprisingly, segregation levels are also higher in urban areas where the occupational status of employed immigrants is more similar to that of natives, i.e., immigrant-native inequality is smaller. The size of the agricultural, industrial, and service sectors is not consistently related to segregation.

\clearpage
\thispagestyle{empty}
\clearpage

\textbf{Panel C: Housing market}. Conditions in local housing markets are also systematically related to segregation. FUAs with high apartment prices consistently have lower levels of segregation, while a higher proportion of households overburdened with housing expenses (> 40\% of income) is not systematically linked to segregation. A higher proportion of homeowners and of households that have lived in the same residence the year before are linked to lower segregation. This is consistent with the inhibition of segregation processes by low housing turnover. Tighter housing markets – characterized by fewer dwellings per resident and a higher occupancy rate – are also robustly associated with higher segregation levels. More recently built housing has a weak positive association with segregation, but does not pass our more conservative robustness tests.

\textbf{Panel D: Flow and composition of immigrant population}. Segregation is higher in FUAs with larger recent inflows of immigrants, but these associations are robust only when considering inflows from the previous year rather than longer time horizons. This indicates higher levels of segregation in popular international migration-destination hubs. Concerning the composition of the immigrant population by origin, we find no robust link between segregation and the proportion of immigrants from MENA or Latin American countries. However, segregation is consistently higher in FUAs with a lower proportion of EU and other ‘Western’ immigrants, a higher proportion of immigrants from Sub-Saharan Africa and Asia, and a higher proportion from lower- and lower-middle-income countries.

\textbf{Panel E: Country characteristics}. Finally, to study links between country characteristics and segregation, we switch from models with country fixed effects to random-intercept multilevel SCAs. We retain the five basic FUA controls, up to four additional FUA covariates, and up to three country-level covariates, for a total of 6,174 model specifications. We focus on indicators of policy and migration history, which operate at the country rather than the urban area level. Panel E of Figure \ref{fig:fig4} finds that neither levels of social expenditure nor the migration and immigrant integration policies captured by the MIPEX index are consistently linked to segregation levels. By contrast, housing policy is associated with segregation: segregation is higher in countries with higher social housing expenditure and a higher proportion of the population receiving housing allowances. However, we do not have data on the degree to which the social housing sector is integrated into or separated from the private housing market. This may crucially shape the impact of social housing on segregation. Finally, segregation is higher in established migration destinations, i.e., countries in which the proportion of immigrants among the total population exceeded 10\% already in 2000. Table \ref{tab:1} summarizes the results of the specification curve analysis.

\textbf{Other dimensions and measures of segregation}. Our findings are remarkably consistent across several alternative operationalisations of segregation (Figures \ref{fig:c1_p}, \ref{fig:c2_5000}, \ref{fig:c3_oth}). When capturing segregation with the isolation rather than the dissimilarity index, associations remain qualitatively similar: the direction of associations and the results of the statistical tests do not change; only the share of Sub-Saharan immigrants and social housing expenditure are no longer robustly associated with segregation. However, most coefficients shrink considerably in size due to the extraordinarily strong dependence of the isolation index on the share of immigrants in the local population (Figure \ref{fig:c1_p} and Methods section). 

Most associations also persist when measuring segregation at the 5km instead of the 1km scale (Figure \ref{fig:c2_5000}). The only notable change is that, at a larger spatial scale, educational composition is no longer related to segregation. This may indicate that highly educated residents are more likely to live in mixed neighbourhoods (e.g., in the city centre), but there is no association with mixing across urban and suburban parts. Associations also persist when we focus on non-EU immigrants rather than all immigrants (Figure \ref{fig:c3_oth}), though the link between country-level social housing expenditure disappears and the MIPEX score becomes more robustly positively linked to segregation. In three additional robustness checks, we examine whether results change when we exclude FUAs with fewer than 100,000 residents (Figure \ref{fig:c4_large}), FUAs with an immigrant population of less than 5\% (Figure \ref{fig:c5_foreign}), and FUAs next to national borders (Figure \ref{fig:c6_borders}). Overall patterns are very similar to the main analysis. However, a recurring observation across robustness checks is that not all indicators of housing market tightness and turnover are as clearly linked to segregation as in the main analysis, although substantive insights remain unchanged. At the country level, social housing expenditure proves to be least robustly associated with segregation across these additional analyses.

\begin{table}[t]
\begin{small}
\setlist[itemize]{topsep=3pt, partopsep=0pt, itemsep=0pt, leftmargin=*}
\centering
\caption{Summary of associations of structural urban area/country characteristics with segregation}
\label{tab:1}
\renewcommand{\arraystretch}{1.1}
\begin{tabular}{>{\centering\arraybackslash}m{0.18\textwidth}m{0.24\textwidth}m{0.22\textwidth}m{0.26\textwidth}}
\textbf{} & \multicolumn{1}{c}{\textbf{Negative}}  & \multicolumn{1}{c}{\textbf{None}} & \multicolumn{1}{c}{\textbf{Positive}} \\
\hline
\textbf{Urban population}
 &
\begin{itemize}
  \item[-] Population density
  \item[-] \% immigrants
  \item[-] \% with higher education
\end{itemize} &
 &
\begin{itemize}
  \item[-] Population dispersion
  \item[-] Population size
  \item[-] \% far-right voters
\end{itemize} \\
\hline
\textbf{Economy} &
\begin{itemize}
  \item[-] Economic \newline performance
  \item[-] Immigrants’ relative unemployment
\end{itemize} &
\begin{itemize}
  \item[-] Sectoral composition
\end{itemize} &
\begin{itemize}
  \item[-] Immigrants’ relative occupational status
\end{itemize} \\
\hline
\textbf{Housing Market} &
\begin{itemize}
  \item[-] \% same residence
  \item[-] \% homeowners
\end{itemize} &
\begin{itemize}
  \item[-] Recency of housing stock
\end{itemize} &
\begin{itemize}
  \item[-] Housing market tightness
  \item[-] Housing affordability
\end{itemize} \\
\hline
\textbf{Migrant flow/ composition} &
\begin{itemize}
  \item[-] \% origin in Western or EU country among immigrants
\end{itemize} &
\begin{itemize}
  \item[-] \% origin in MENA and Latin America among immigrants
\end{itemize} &
\begin{itemize}
  \item[-] Recent international inflow
  \item[-] \% origin in low-income countries, Sub-Saharan Africa, or Asia among immigrants
\end{itemize} \\
\hline
\textbf{Country Characteristics} &
 &
\begin{itemize}[leftmargin=*]
  \item[-] Social expenditure
  \item[-] Migration policy
\end{itemize} &
\begin{itemize}[leftmargin=*]
  \item[-] Established destination country
  \item[-] Social housing expenditure / allowances
\end{itemize} \\\hline
\end{tabular}
\end{small}
\end{table}

\section{Discussion}

This study provides the most comprehensive assessment of immigrant-native residential segregation across Europe, covering all 717 Functional Urban Areas (FUAs) in 30 countries. Our findings reveal that segregation is higher in Western and Northern Europe vis-à-vis large parts of Eastern and Southern Europe. However, 60\% of the variation in segregation levels occurs within countries. We find that segregation is systematically linked to urban area population characteristics, economy, housing market, immigrant inflow and composition. At the country level, segregation is linked to migration histories and housing policies, but unrelated to migration policy.

Our study makes two key contributions. First, the results expand, update, and systematize our knowledge of immigrant-native segregation across Europe. Our study provides harmonised, high-resolution segregation estimates for all urban areas in 30 European countries. While such large-scale comparisons of segregation levels across urban areas have been carried out in the US \cite{Iceland.2008, Lichter.2024, Lichter.2015, Parisi.2025, Reardon.2008}, equivalent evidence in Europe has so far been missing.  We show that immigrant-native segregation is highly uneven across Europe, with a clear North/West vs. South/East divide – although Poland's high levels of unevenness are somewhat outstanding. Moreover, non-EU immigrants are typically more segregated than their EU counterparts, with urban areas near EU borders representing a notable exception. We also show that segregation in European urban areas is strongly shaped by large-scale patterns, with diverse urban cores contrasting sharply with the underrepresentation of immigrant minorities in suburban areas and commuting zones. 

Second, we examine a broad set of urban and national characteristics as potential correlates of segregation. As summarised in Table \ref{tab:1}, segregation is consistently higher in urban areas with larger populations and greater spatial dispersion of the population, but lower in urban areas with high population density and larger immigrant populations. Segregation is also lower in urban areas with a more educated population, fewer right-wing voters, higher economic performance, and lower economic inequality. Associations with economic immigrant-native inequality are less consistent: disproportionately high unemployment rates among immigrants are related to higher segregation, but a relatively higher occupational status of immigrants is also related to higher, not lower segregation. More housing-turnover and greater tightness of the housing market are linked to higher segregation – as are immigrant populations from lower-income countries, Sub-Saharan Africa, and Asia, rather than from the EU or other `Western' countries. International immigration hubs that have recently attracted many immigrants are more segregated. Across countries, however, more established immigrant destinations are more segregated, indicating a local consolidation of immigrant groups in these contexts. Segregation is also stronger in countries with greater social housing expenditure and greater uptake of housing allowances, while migration and integration policies are not connected to segregation levels. 

Several findings align with long-standing theoretical expectations of segregation. For example, the link between higher housing-turnover and greater segregation is consistent with Schelling's (1978) segregation model, which posits that segregation emerges through repeated residential moves. Similarly, our findings align with place stratification theory \cite{Logan.1993}, highlighting the role of discrimination: segregation is higher in areas with stronger support for far-right parties and in cities with larger shares of stigmatised origin-groups. While overall economic inequality is linked to higher segregation, our findings on native–immigrant inequality are more nuanced than previous findings \cite{Fossett.2011, Iceland.2006}: segregation is higher when unemployment gaps are larger, but also when occupational differences among the employed are smaller – supporting spatial assimilation theory in the first case, and yet contradicting it in the second. Finally, higher segregation levels in major immigration destinations and countries with a long-standing immigration legacy underscore the role of ethnic enclaves and path-dependent settlement patterns (even more so for non-EU immigrants).

At the same time, we identify systematic links between segregation levels and structural characteristics that segregation theories remain largely silent about. We find that dispersed settlement structures facilitate separation, highlighting the role of metropolitan morphology as a macro-spatial feature of segregation. Higher education levels are associated with more mixing on the small-scale neighbourhood level, but not at the macro scale between urban cores and suburban surroundings. FUAs with higher apartment prices exhibit lower segregation levels, possibly because high housing costs constrain households' ability to act on discriminatory preferences, forcing them to prioritize affordability over selective location choices \cite{Logan.1993}. This interpretation is further supported by our finding that high housing prices reduce segregation more among non-EU immigrants, toward whom exclusionary preferences may be more pronounced. Concerning policy, earlier studies have suggested that countries with multiculturalist approaches exhibit lower levels of segregation \cite{Koopmans.2010}. However, drawing on a larger and more comprehensive cross-country dataset than previous research, the association between migration policy and segregation does not appear robust. We instead find that higher social housing expenditure and greater uptake of housing allowances are linked to higher levels of segregation. However, social housing systems differ substantially in their design – some are closely integrated with the private housing market, while others are more separated – shaping the extent to which these policies influence residential segregation \cite{Andersson.2018, Marcinczak.2023}.

We acknowledge three main limitations of the present study. First, a drawback of our broad geographical scope and the harmonised data is that information on the immigrant population is less granular. Accordingly, we cannot differentiate immigrants by country of origin, length of residence, or socio-economic status. We also cannot compare the residential patterns of immigrants to those of their descendants across Europe. However, a robustness check (Figure \ref{fig:d_nl}) with more detailed data from the Netherlands shows a strong correlation (r = 0.9) of segregation levels among immigrants and their descendants. Still, documenting the variation across generations can be important for understanding long-term integration trajectories.

Second, while the 1$\times$1 km grid resolution of the census data ensures comparability across Europe, it does not capture geographical sorting at smaller spatial scales, such as residential blocks or streets. Such micro-level segregation within census grid cells may be particularly pronounced in densely populated urban areas. At the same time, the neighbourhood within 1km reach seems to be important for many aspects of social integration, such as day-to-day interaction, local amenities, and school catchment areas \cite{Reardon.2004}. 

Third, our analysis of the structural correlates of segregation is observational and cannot identify causal effects. Patterns of segregation emerge over time through complex interactions between individual residential choices, institutional practices, and urban development \cite{Desmond.2019, Elbers.2024, Krysan.2017, Logan.2012, Logan.1993}. Some of the correlates we have considered, such as local inequality between natives and immigrants, can not only be a precursor, but also a consequence of segregation. Instead of a causal assessment, the Specification Curve Analysis provides a rich descriptive overview of the correlates of segregation, accounting for a comprehensive set of potential covariates.

Nevertheless, our documentation of segregation across Europe and the analysis of its macro-level correlates offers several opportunities for future research. First, our harmonised segregation indices can serve as a valuable resource for researchers who need to strategically select urban areas for in-depth studies, such as qualitative or longitudinal approaches. Such studies can explore how local institutions, histories, and policies interact to shape segregation dynamics over time \cite{Bruch.2019}. Second, our comprehensive analysis of the correlates of segregation provides a starting point for future causal research investigating whether the associations persist in longitudinal designs and how immigrant-native segregation co-evolves with other dimensions of segregation, such as socio-economic segregation \cite{Ubareviciene.2025}. As such, our findings provide a foundation for refining and expanding theoretical frameworks on the contextual drivers of segregation. Finally, our documentation of substantial variation in segregation levels across and within countries raises the question of whether differences in local segregation levels translate into diverging lived experiences among immigrants and whether they are mirrored in their integration outcomes. Accordingly, our findings also call for further research on the consequences of residential segregation.

\clearpage

\section*{Methods} \label{sec:methods}

\subsection*{Data}

\textbf{Functional Urban Areas}. To identify urban areas across Europe, we rely on the joint Eurostat/OECD classification of Functional Urban Areas (FUAs). Each FUA consists of an urban centre/city and its commuting zone, i.e., all surrounding areas in which at least 15\% of the population commute to the urban centre \cite[for details, see][]{Eurostat.2025a}. This uniform way of identifying urban areas across all 30 countries avoids measurement issues arising from country-specific variations in administrative definitions of urban areas and has become standard in the comparative European literature on segregation \cite{Marcinczak.2023, Benassi.2020}. The definition of the FUAs also closely resembles the definition of metropolitan areas usually used in the US segregation literature \cite{Lichter.2015, Reardon.2008}. We use shapefile information on the geographical boundaries of FUAs provided by Eurostat.

\textbf{Local native and immigrant population}. To identify the spatial distribution of the local native and immigrant population, we rely on data from the 2021/2022 censuses conducted in all 30 countries. For all countries but the UK, Eurostat provides aggregated and harmonized data. For all 1$\times$1 km grid cells, these data capture the number of residents born in the country of residence, in another EU country, and in a non-EU country \cite{Eurostat.2025}. For the United Kingdom, we harmonised data from three separate sources: The 2021 census for England and Wales, the 2022 census for Scotland, and the 2021 census for Northern Ireland. While the latter is available in a 1$\times$1 km grid cell format directly comparable to the remaining European data, the other two sources provide the relevant information at the spatial resolution of Lower Statistical Output Areas (LSOA, 1,400 inhabitants on average). We performed dasymetric mapping techniques to translate those into a 1$\times$1 km regular grid. Intuitively, we first used the data from the input layer and distribute the characteristics of each LSOA's along a 100m resolution population layer. In this step, the population characteristics are distributed to match the the number of residents along the 100m resolution population layer. Subsequently, we interpolated from this 100m layer to the final 1$\times$1 km grid. This provides us with a harmonized set of 1$\times$1 km grid cells and their composition in terms of natives and immigrants across all 30 countries.

\textbf{FUA and country-level correlates of segregation}. To examine the correlates of residential segregation, we collect FUA- and country-level information from Eurostat, the OECD, the UN, the European Election Database, and the Migrant Integration Policy Database. Missing values were imputed using multivariate imputation by chained equations with 20 imputed datasets \cite{Buuren.2011}. For each variable, Table \ref{tab:variables} in the Supplementary Material lists the definition, the spatial level of measurement, and the year of measurement. 

\subsection*{Measures}

Our primary measure of segregation captures the unevenness of the distribution of immigrants and natives across the urban area, i.e., the degree to which both groups live spatially separated from one another \cite{Massey.1988, Reardon.2004}. We measure unevenness at the FUA level with the spatial dissimilarity index $\widetilde{D}$. Compared to the non-spatial version of the dissimilarity index $D$, the spatial index $\widetilde{D}$ takes the geographic locality of the grid cells into account by interpolating population characteristics within a given radius \cite{Reardon.2008, Reardon.2004}. It is defined as:
\begin{equation}
 \widetilde{D}=\sum_{m=1}^{M}\int_{p\in R}{\frac{\tau_p}{2TI}\left|{\widetilde{\pi}}_{pm}-\pi_m\right|dp},
\end{equation}
where $\tau_p$ denotes the population density at location $p$, ${\widetilde{\pi}}_{pm}$ is the local proportion of group m around point $p$, and $\pi_m$ is the overall proportion of group m in the city. The term $T$ represents the total population, and $I=\sum_{m=1}^{M}{\pi_m\left(1-\pi_m\right)}$ is a normalization constant. Intuitively, $\widetilde{D}$ aggregates the deviation of the local immigrant-native composition of all 1$\times$1km grid cells in a FUA from the FUA’s mean immigrant-native composition weighted it by local population density. Higher values therefore imply that the local immigrant-native compositions deviate more strongly from the mean composition of the FUA, i.e., that immigrants and natives live more segregated from each another. We prefer the spatial dissimilarity index $\widetilde{D}$ over other measures of unevenness, as its two-group measure is scale-invariant to population size \cite{Reardon.2004}, ensuring comparability across urban areas and countries with differently sized population. For our main analysis, we use a radius of 1km to define the local neighbourhood, as this approximates institutional neighbourhoods within which most individuals can reach essential amenities \cite{Reardon.2008}. We also provide supplementary results for other scales, such as 500m, 2km, and 5km. 

While our main analysis focuses on unevenness, we also provide additional analyses on isolation as a second key dimension of segregation \cite{Massey.1988, Reardon.2004}. The spatial isolation index ${_m\widetilde{P}}^\ast$ for group captures the extent to which immigrants $m$ are exposed to immigrants rather than natives in their local environments. It is defined as:
\begin{equation}
{_m\widetilde{P}}^\ast=\int_{p\in R\ }\frac{\tau_{pm}}{T_m}{\widetilde{\pi}}_{pm}dp,
\end{equation}
where $\tau_{pm}$ denotes the population density of group $m$ for the local grid cell $p$, $T_m$ is the total population of group $m$, and ${\widetilde{\pi}}_{pm}=\frac{{\widetilde{\tau}}_{pm}}{{\widetilde{\tau}}_p}$ corresponds to the proportion of group $m$ in the local neighbourhood around grid cell $p$. Intuitively, the isolation index quantifies the average proportion of immigrants in immigrants' local environment. In addition to the spatial distribution of natives and immigrants, it is thus highly dependent on the overall share of immigrants in an urban area. All spatial segregation indices were calculated using the R package seg \cite{Hong.2014}.

\subsection*{Statistical Methods}

To investigate which FUA- and country characteristics are robustly associated with residential segregation, we use specification curve analysis for inference with non-experimental data \cite{Simonsohn.2020}. We first define the universe of feasible specifications that are also plausible \cite{Auspurg.2025}. 

For all urban area-level analysis, we run regression models with country fixed effects. All models include basic population controls (population size, population density, population dispersion, percentage of immigrants, educational composition). In addition, they contain a set of one to four additional FUA-level variables. These sets include all possible specifications, excluding those combinations with multiple variables from within the same variable subgroup. For instance, both GDP per capita and the unemployment rate capture economic performance, so they are not included in the same model. Table \ref{tab:variables} in Supplementary Material provides an overview of all variable subgroups and the combinations of variables that are (not) estimated in the same model. These criteria result in a total of 9,990 fixed-effects models for the urban area-level covariates.

For the analysis of country-level characteristics, we use random-effects multilevel models with random intercepts. Next to FUA-level population controls and up to four additional FUA characteristics, we include up to three country-level characteristics in these models. This results in a total of 6,174 random-effects multilevel models. 

For each variable $v$, the specification analysis produces a distribution of $K_v$ point estimates for the association between $v$ and the segregation score $y$ across specifications $k=1,\ldots K_v$. We then perform three tests of statistical significance of each variable $v$, as illustrated in Figure \ref{fig:fig3} \cite{Simonsohn.2020}. In Test 1, our least conservative test, we assess whether at least 95\% of the $K_v$ estimates fall on one side of the null line, indicating a robust association in terms of estimated direction. For Tests 2-3, we compare the observed specification curve to counterfactual specification curves obtained by bootstrapping. For each variable $v$ and each specification $k$, we estimate a counterfactual segregation score $y_{k_v}^\ast$ imposes a null association between the variable $v$ and segregation $y$. We achieve this null association by defining $y_{k_v}^\ast$ as the original segregation score $y$ minus the estimated effect ${\hat{b}}_{K_v}$ of variable $v$ on the segregation score in specification $k$ multiplied with the observed $v$: $y_{k_v}^\ast=y-\ {\hat{b}}_{K_v} v$. All other FUA and country characteristics remain unchanged. From this dataset, we draw samples of the size of the original number of FUAs (717) with replacement and estimate all specifications $K_v$ on those data. Repeating this process 500 times provides us with a bootstrapped distribution of counterfactual estimates under an imposed null association for each variable $v$. Test 2 compares the median estimate from the observed specification curve to the median estimates of the counterfactual specification curve; if the observed median is more extreme than the counterfactual medians in at least 95\% of the bootstrapped distributions, the variable passes Test 2. Test 3 ranks the estimates in both the observed and the counterfactual specification curves by size. For each specification, it then tests whether the observed estimates fall outside of the 95\% quantile range of the corresponding counterfactual estimates. The variable passes Test 3 if the observed estimate falls outside of the 95\% quantile range for at least 95\% of all specifications.

\clearpage

\bibliography{references}

\clearpage
\appendix

\setcounter{table}{0}
\renewcommand{\thetable}{S\arabic{table}}%
\setcounter{figure}{0}
\renewcommand{\thefigure}{S\arabic{figure}}%

\part*{\center Supplementary material} 

This supplement consists of four parts.
\begin{itemize}
\item	Part A contains additional descriptive plots related to the spatial Dissimilarity Index with a 1km radius. It includes further maps illustrating immigrant native distributions in urban areas with different levels of segregation (A1) as well as scatterplots showing the relationship with related dissimilarity-based measures of segregation (A2 and A3).
\item	Part B repeats the regional patterns shown in the density plot in Figure 2 for different measures of segregation (B1). It also provides additional versions of Figure 2 for the spatial dissimilarity index with radii of 500 m (B2) and 5 km (B3), for the spatial isolation index (B4). Finally, the country-specific boxplots showing the spatial dissimilarity indices at different scales are shown (B5).
\item	Part C provides additional iterations of the results of the specification curve analysis shown in Figure 4. The materials compare the original results to the spatial isolation index at a 1km radius (C1), the spatial dissimilarity index at a 5km radius (C2), the spatial dissimilarity index at a 1km radius for non-EU immigrants only (C3), the spatial dissimilarity index at a 1km radius for the subset of FUAs with at least 100,000 inhabitants (C4), with at least 5\% immigrants (C5), which are not located next to national borders (C6). 
\item Part D illustrates the link between the residential locations and segregation indices of immigrants and their descendants (D1).
\item Part E lists the definitions and measures of all variables included in Specification Curve Analysis.
\end{itemize}

\clearpage

\section{Additional Descriptives} \label{suppl:A}

\begin{figure}[h!]
  \centering
  \includegraphics[width=1\linewidth]{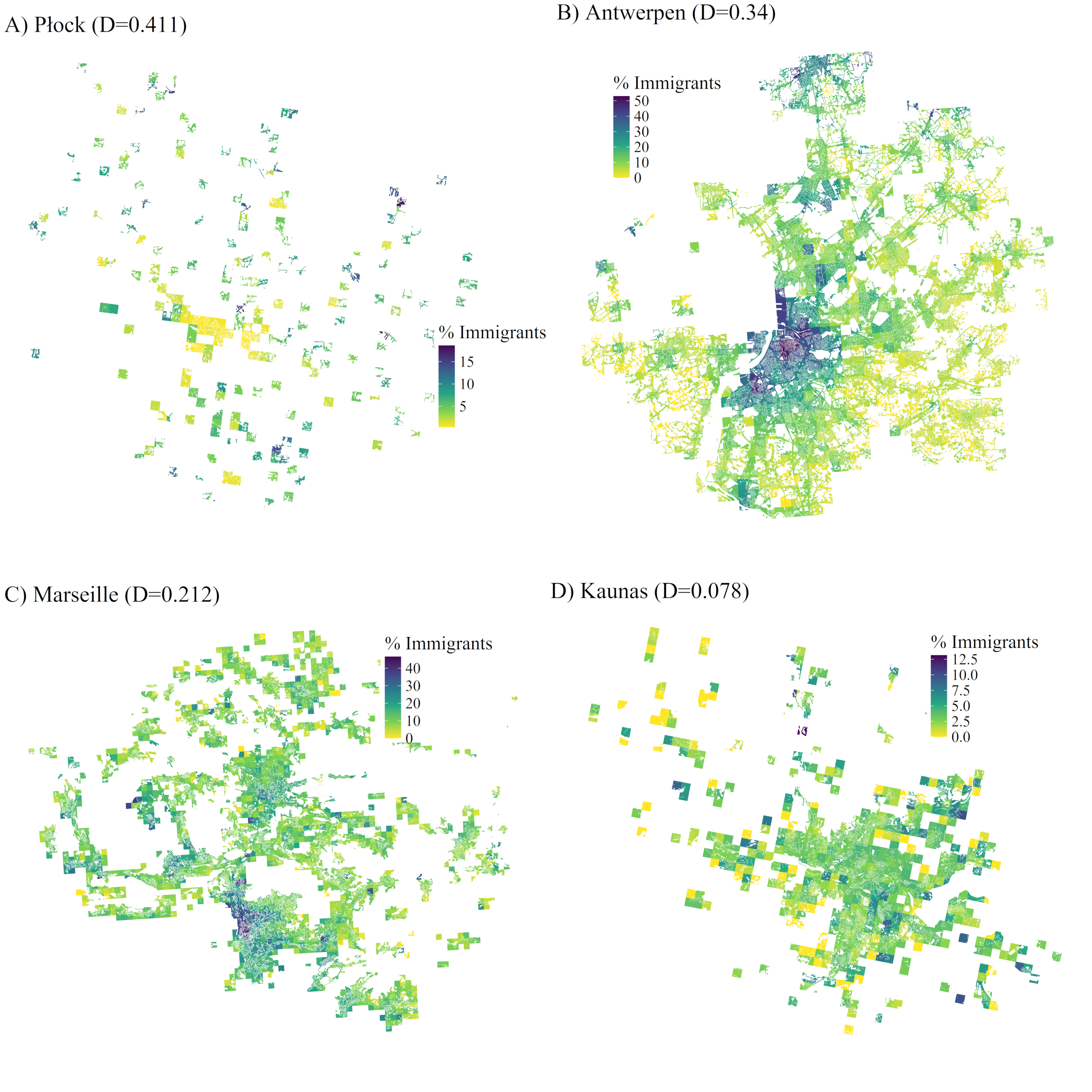}
  \caption{Additional examples of FUAs. The colours indicate the share of immigrants in each 1x1km grid cell.}
  \label{fig:step}
\end{figure}

\clearpage

\begin{figure}[h!]
  \centering
  \includegraphics[width=1\linewidth]{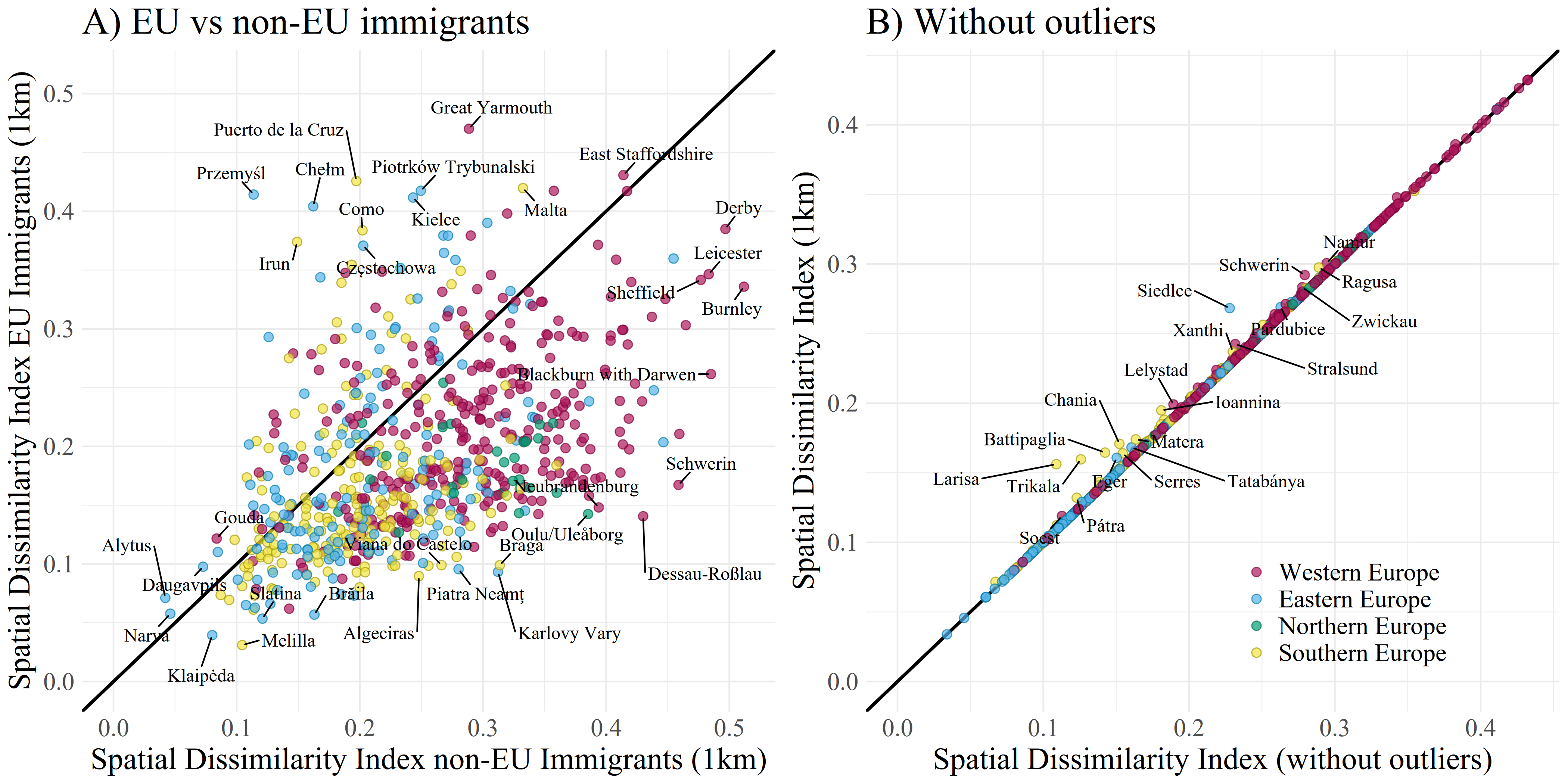}
  \caption{A) Scatterplot of the dissimilarity index (1km) for EU immigrants and non-EU immigrants with 45 degree line; B) Scatterplot of the dissimilarity index (1km) based on the full set of observations and based on a subset without outliers (549 grid cells) with 45 degree line. Outliers were defined as having at least 50\% foreign residents and either a) at least 5-fold share of foreign residents as compared to the 4 nearest grids cell, or b) twice as many male than female residents.}
  \label{fig:degree1}
\end{figure}

\begin{figure}[h!]
  \centering
  \includegraphics[width=1\linewidth]{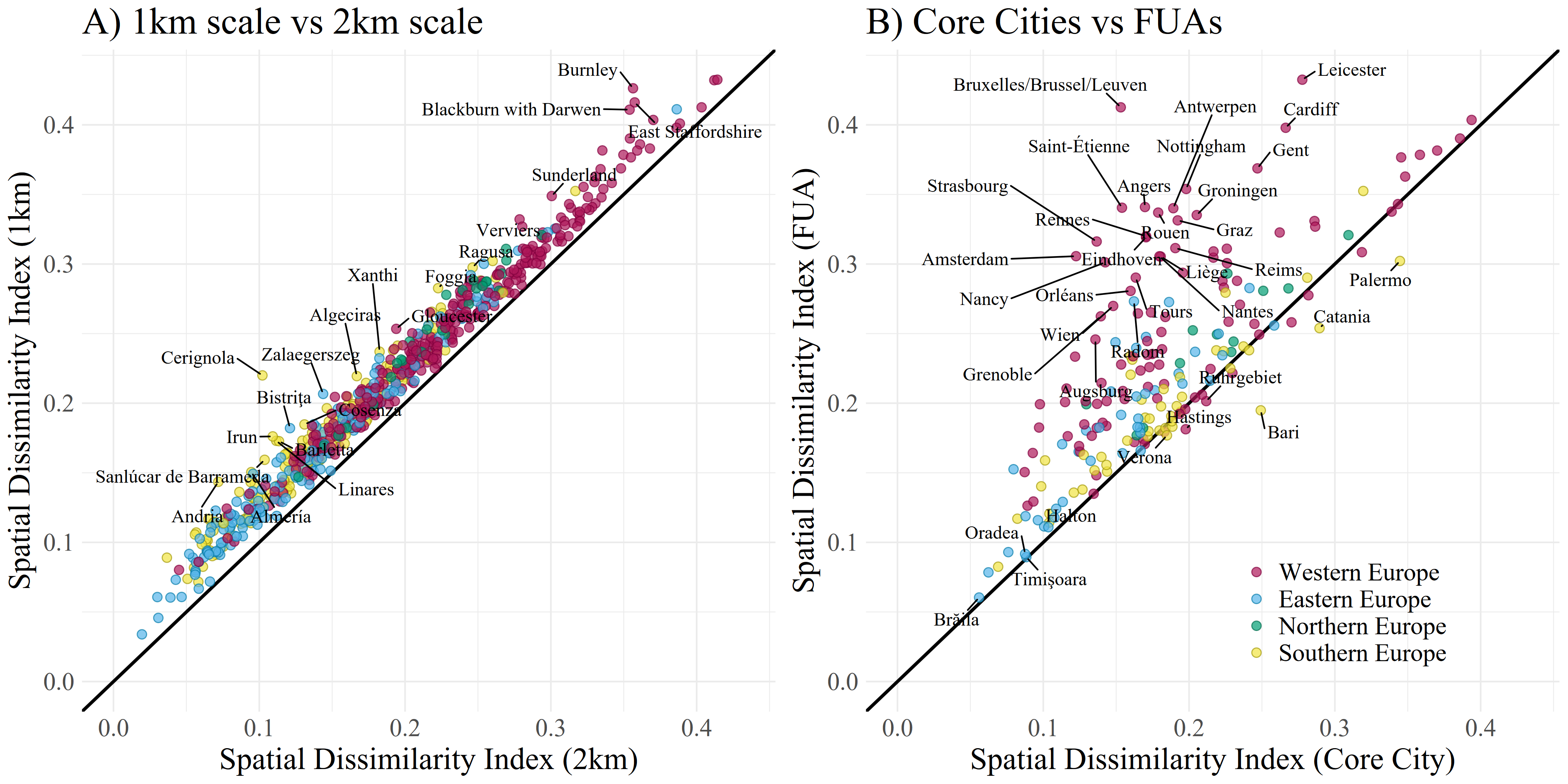}
  \caption{A) Scatterplot of dissimilarity index with 1km and dissimilarity index with 2km bandwidth with 45 degree line; B) Scatterplot of dissimilarity index with 1km for the FUAs and cities with 45 degree line based on a subset of 199 cities with a unique metropolitan city according to OECD definition and spatial boundaries.}
  \label{fig:degree2}
\end{figure}

\clearpage

\section{Different Measures of Segregation} \label{suppl:B}

\begin{figure}[h!]
  \centering
  \includegraphics[width=0.9\linewidth]{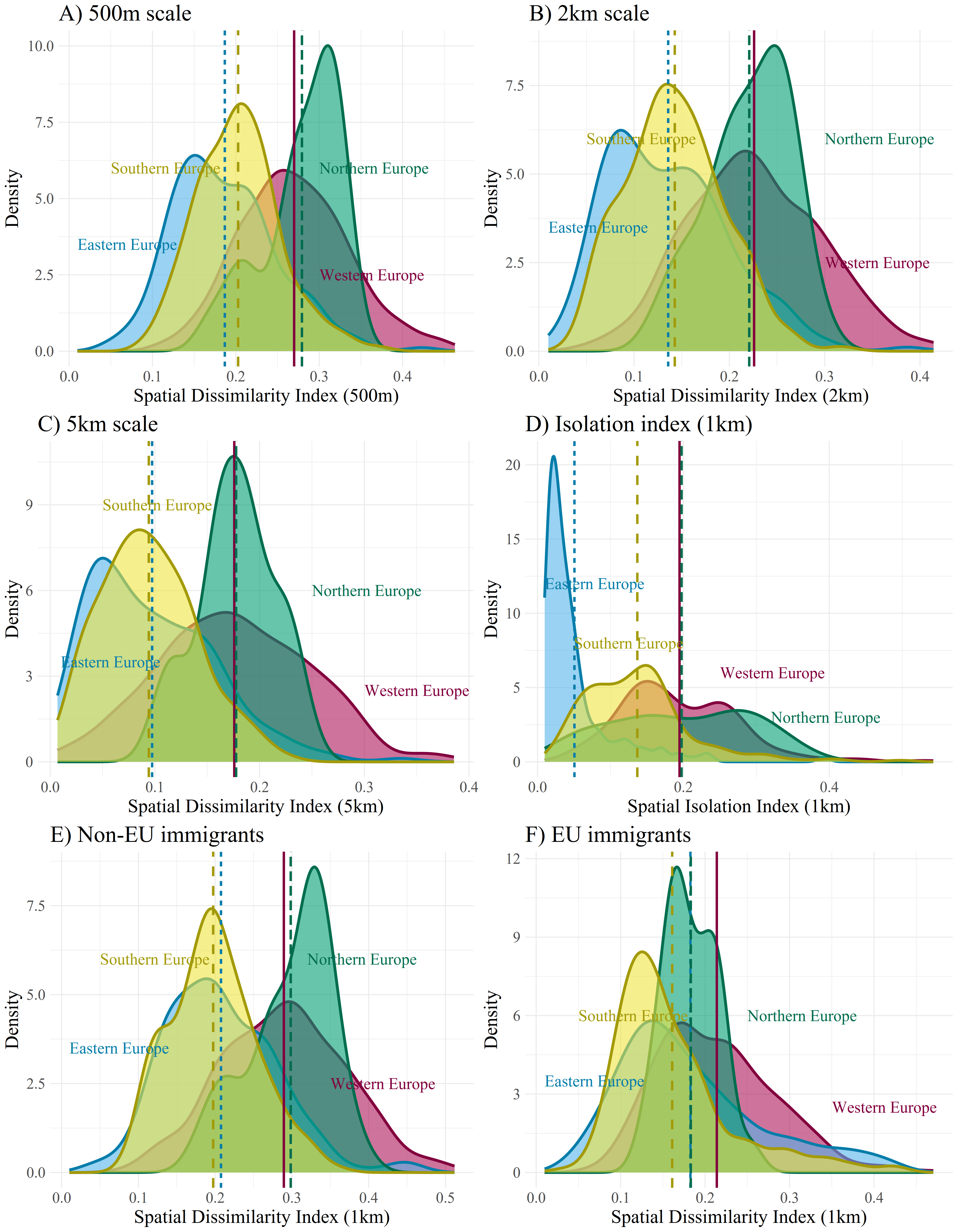}
  \caption{Density distribution of segregation indices across regions for A) 2km scale, B) 5km scale, C) spatial immigrant-isolation index on 1km scale, D) spatial immigrant isolation on 5km scale, E) non-EU immigrants only, and F) EU immigrants only.}
  \label{fig:density}
\end{figure}

\clearpage

\begin{figure}[h!]
  \centering
  \includegraphics[width=1\linewidth]{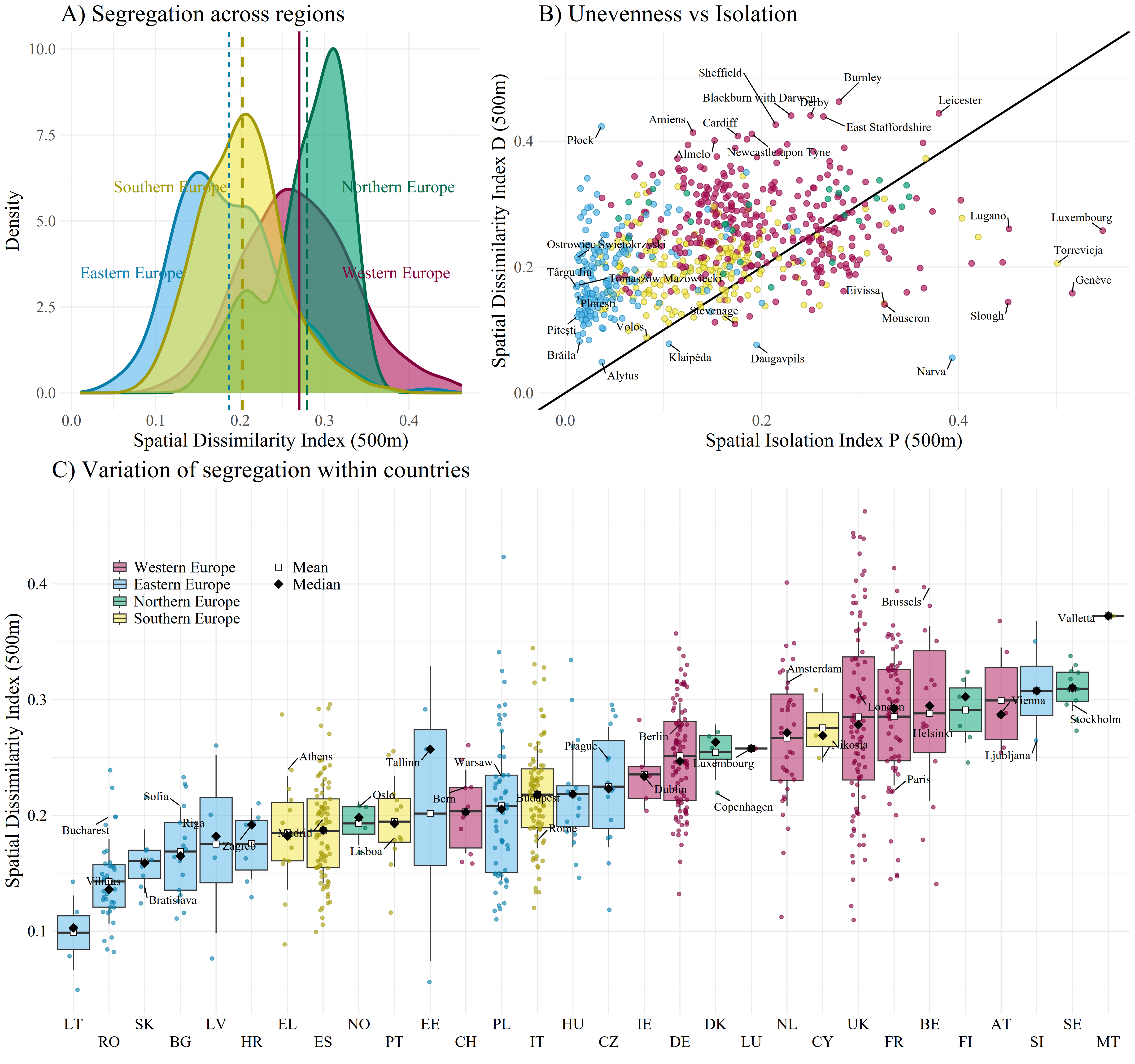}
  \caption{Variation of segregation across Europe for 500m neighbourhood scale. A) Density distribution of segregation indices across regions; B) Comparison of spatial Dissimilarity and spatial Isolation index with 45 degree line;  C) Residential Segregation in 717 FUAs across 30 countries and the FUAs including their capital cities. Note: The box of each boxplot represents the interquartile range (IQR) across FUAs and the whiskers extend to the smallest and largest values within 1.5xIQR. Countries are ordered by the mean of the segregation index across their FUAs. 717 FUAs. }
  \label{fig:fig2_500m}
\end{figure}

\clearpage

\begin{figure}[h!]
  \centering
  \includegraphics[width=1\linewidth]{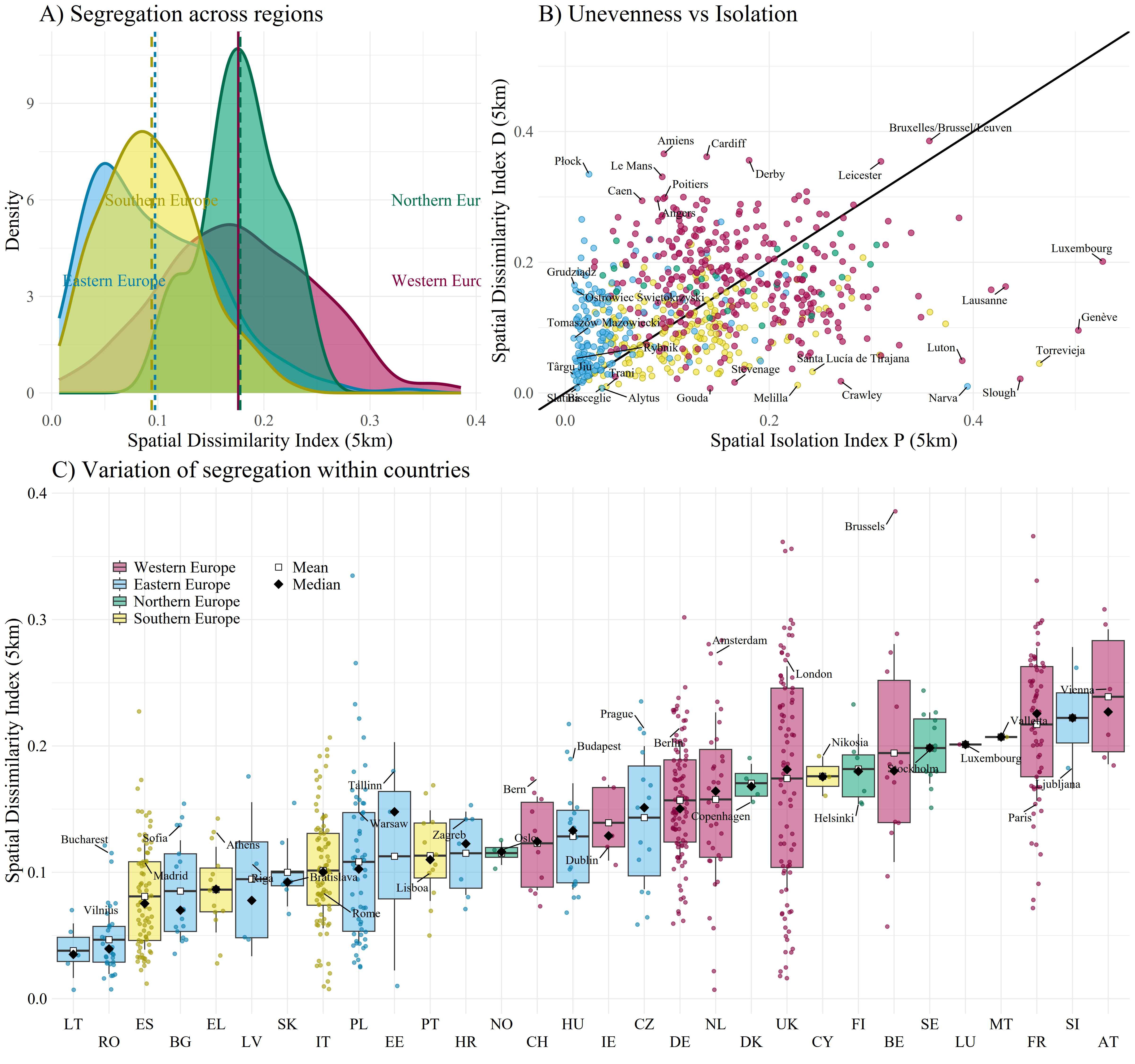}
  \caption{Variation of segregation across Europe for 5km neighbourhood scale. A) Density distribution of segregation indices across regions; B) Comparison of spatial Dissimilarity and spatial Isolation index with 45 degree line;  C) Residential Segregation in 717 FUAs across 30 countries and the FUAs including their capital cities. Note: The box of each boxplot represents the interquartile range (IQR) across FUAs and the whiskers extend to the smallest and largest values within 1.5xIQR. Countries are ordered by the mean of the segregation index across their FUAs. 717 FUAs.}
  \label{fig:fig2_5km}
\end{figure}

\clearpage

\begin{figure}[h!]
  \centering
  \includegraphics[width=1\linewidth]{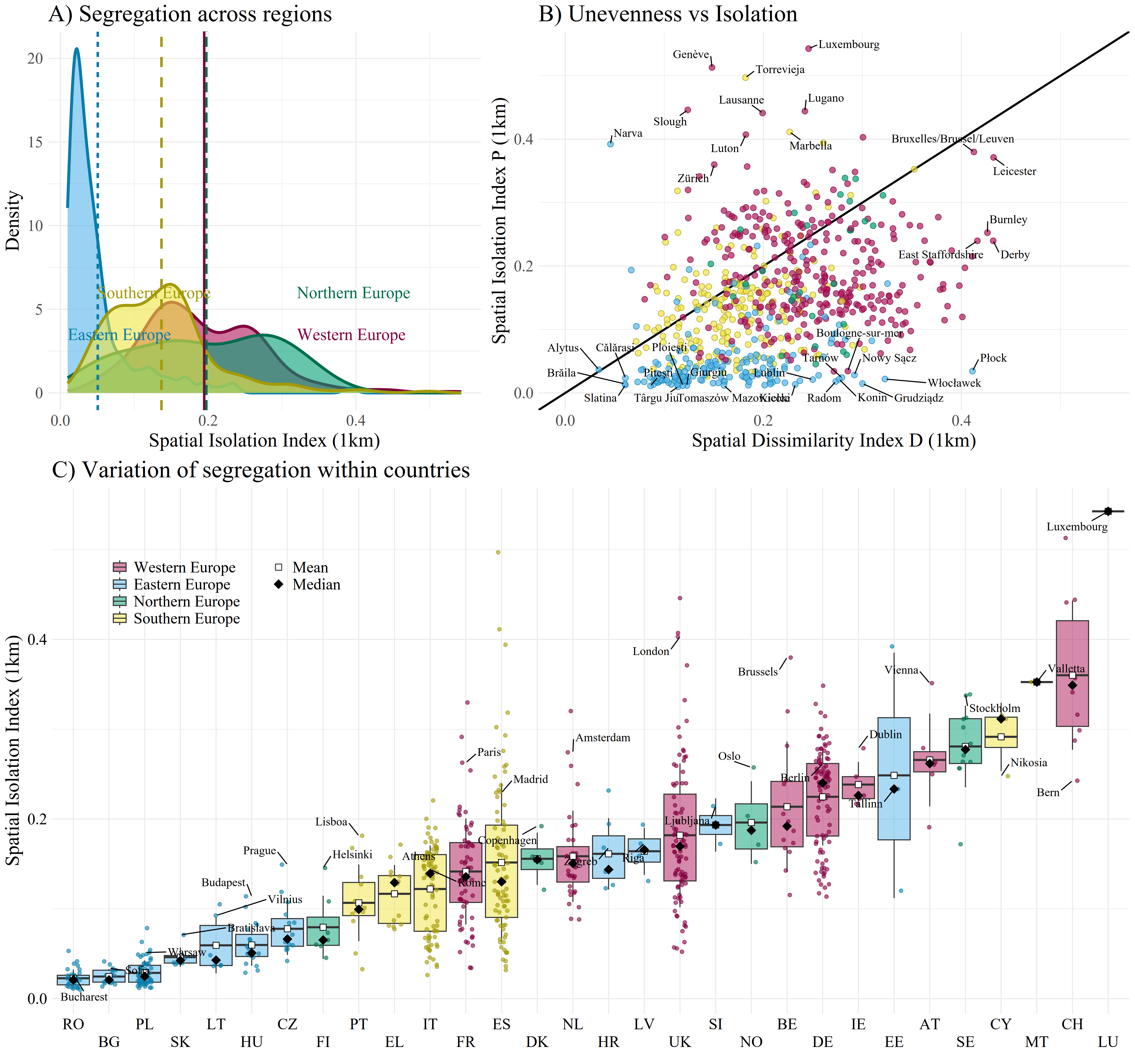}
  \caption{Variation of segregation measured as spatial (immigrant-)Isolation across Europe. A) Density distribution of segregation indices across regions; B) Comparison of spatial Dissimilarity and spatial Isolation index with 45 degree line;  C) Residential Segregation in 717 FUAs across 30 countries and the FUAs including their capital cities. Note: The box of each boxplot represents the interquartile range (IQR) across FUAs and the whiskers extend to the smallest and largest values within 1.5xIQR. Countries are ordered by the mean of the segregation index across their FUAs. 717 FUAs. }
  \label{fig:fig2_p}
\end{figure}

\clearpage

\begin{figure}[h!]
  \centering
  \includegraphics[width=1\linewidth]{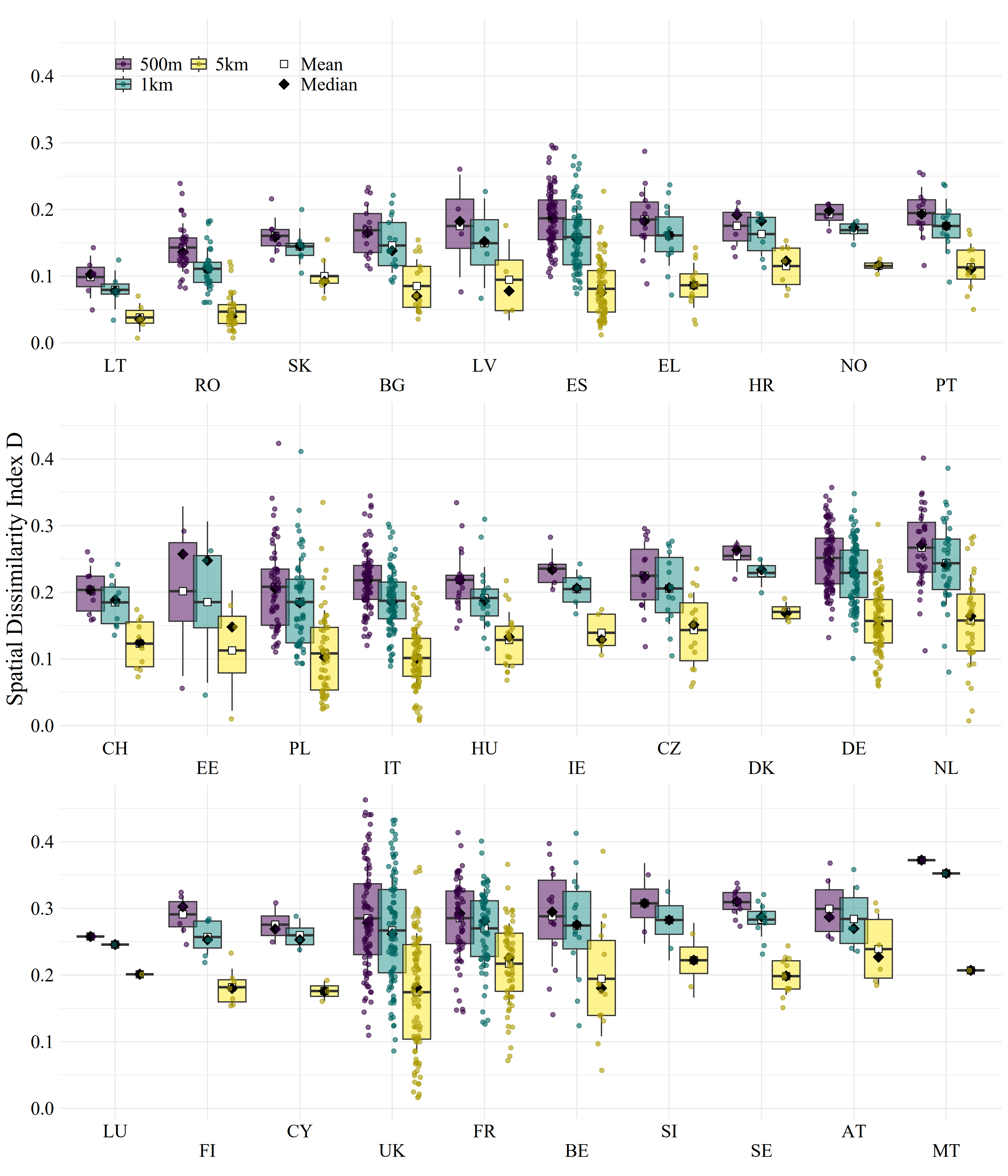}
  \caption{Comparison of variation of segregation for 500m, 1km, and  5km neighbourhood scale in 717 FUAs across 30 countries and the FUAs. Note: The box of each boxplot represents the interquartile range (IQR) of INS across FUAs and the whiskers extend to the smallest and largest values within 1.5xIQR. Countries are ordered (from top left to bottom right) by the mean of the segregation index for 1km scale across their FUAs.}
  \label{fig:scale_box}
\end{figure}

\section{Alternative results for SCA} \label{suppl:C}

\begin{figure}[h!]
  \centering
  \includegraphics[width=0.90\linewidth]{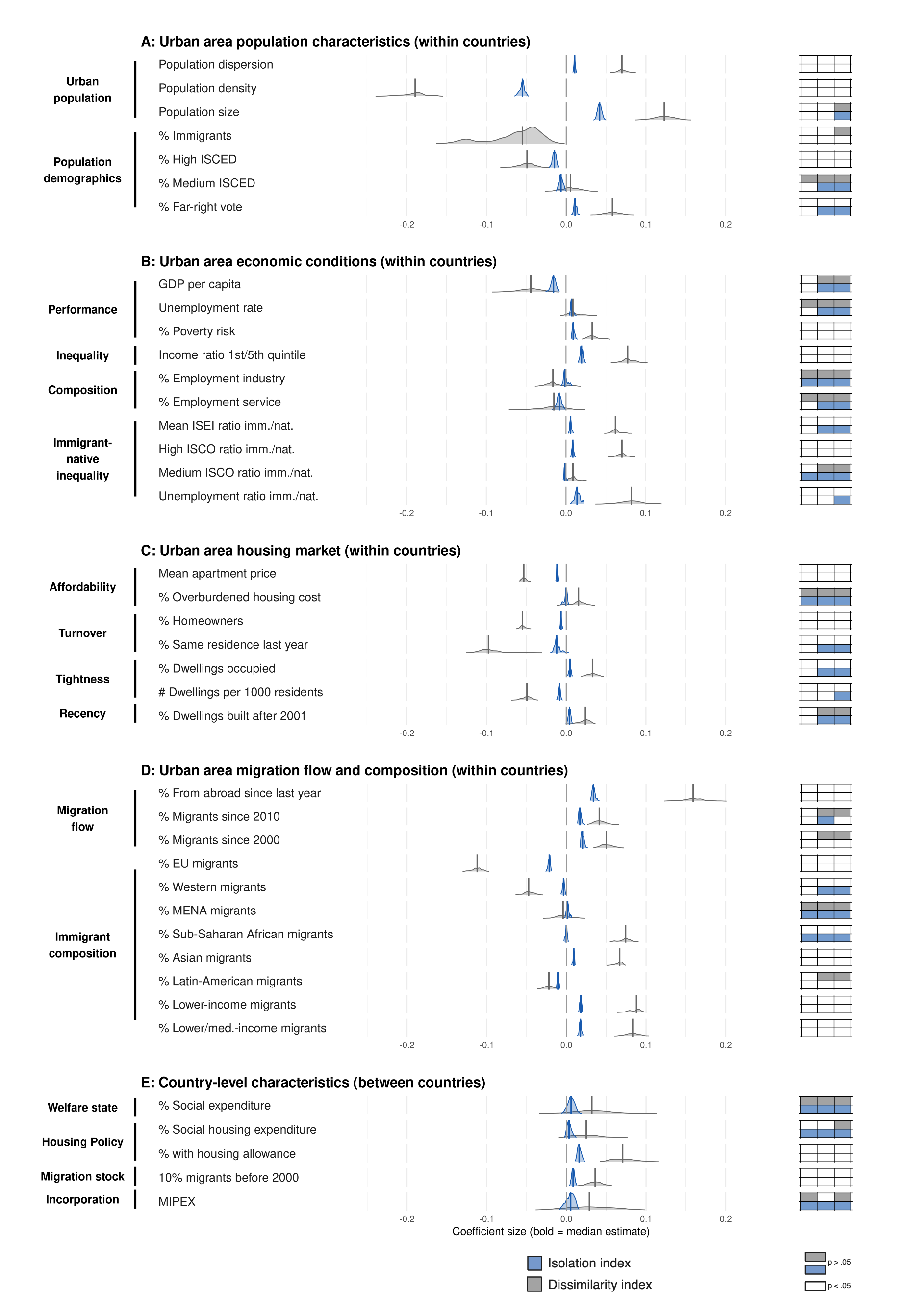}
  \caption{Results for isolation index (blue) vs. dissimilarity index (grey).}
  \label{fig:c1_p}
\end{figure}

\begin{figure}[h!]
  \centering
  \includegraphics[width=0.95\linewidth]{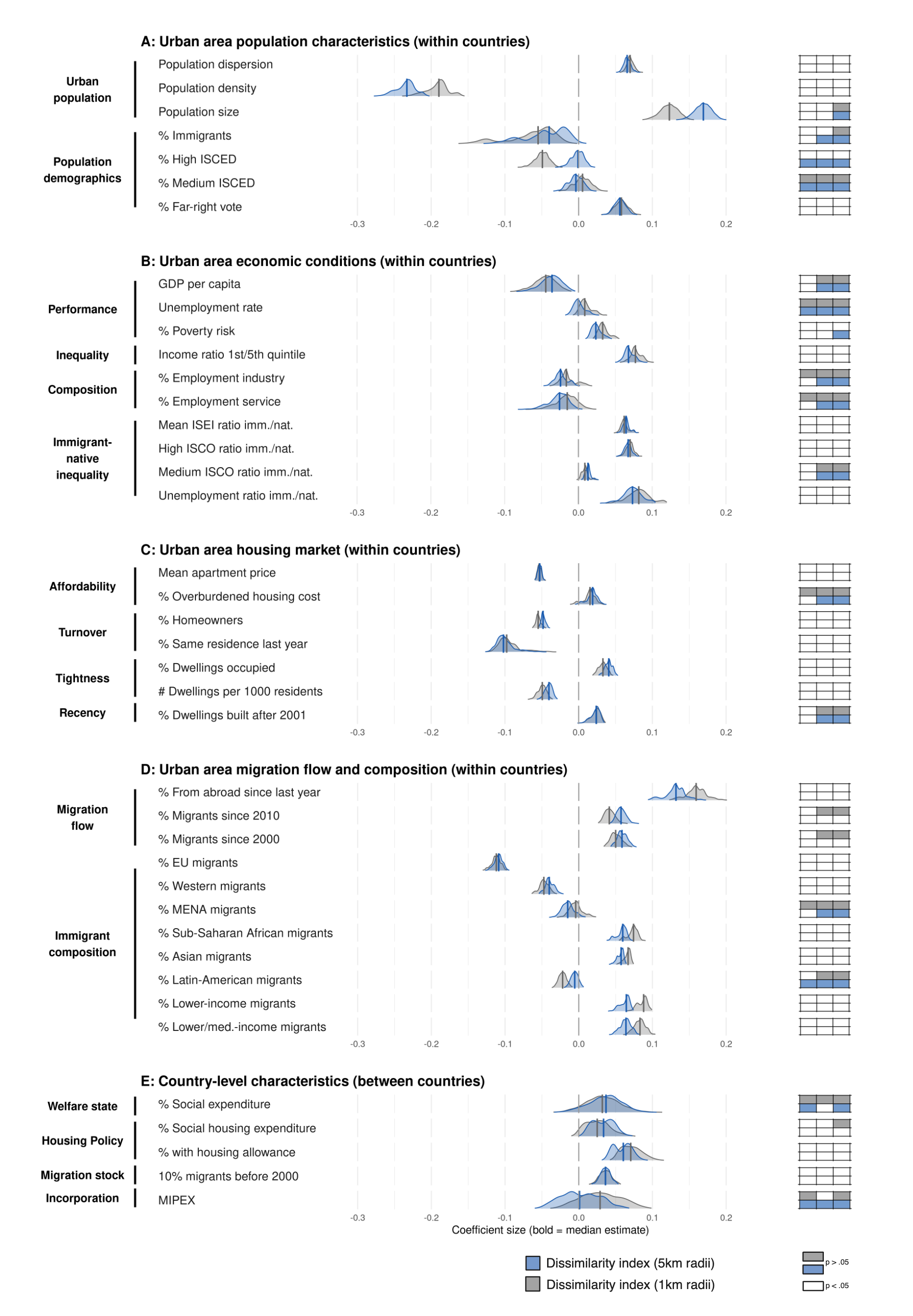}
  \caption{Results for dissimilarity index with 5km (blue) vs. 1km (grey) neighbourhoods radii.}
  \label{fig:c2_5000}
\end{figure}

\begin{figure}[h!]
  \centering
  \includegraphics[width=0.95\linewidth]{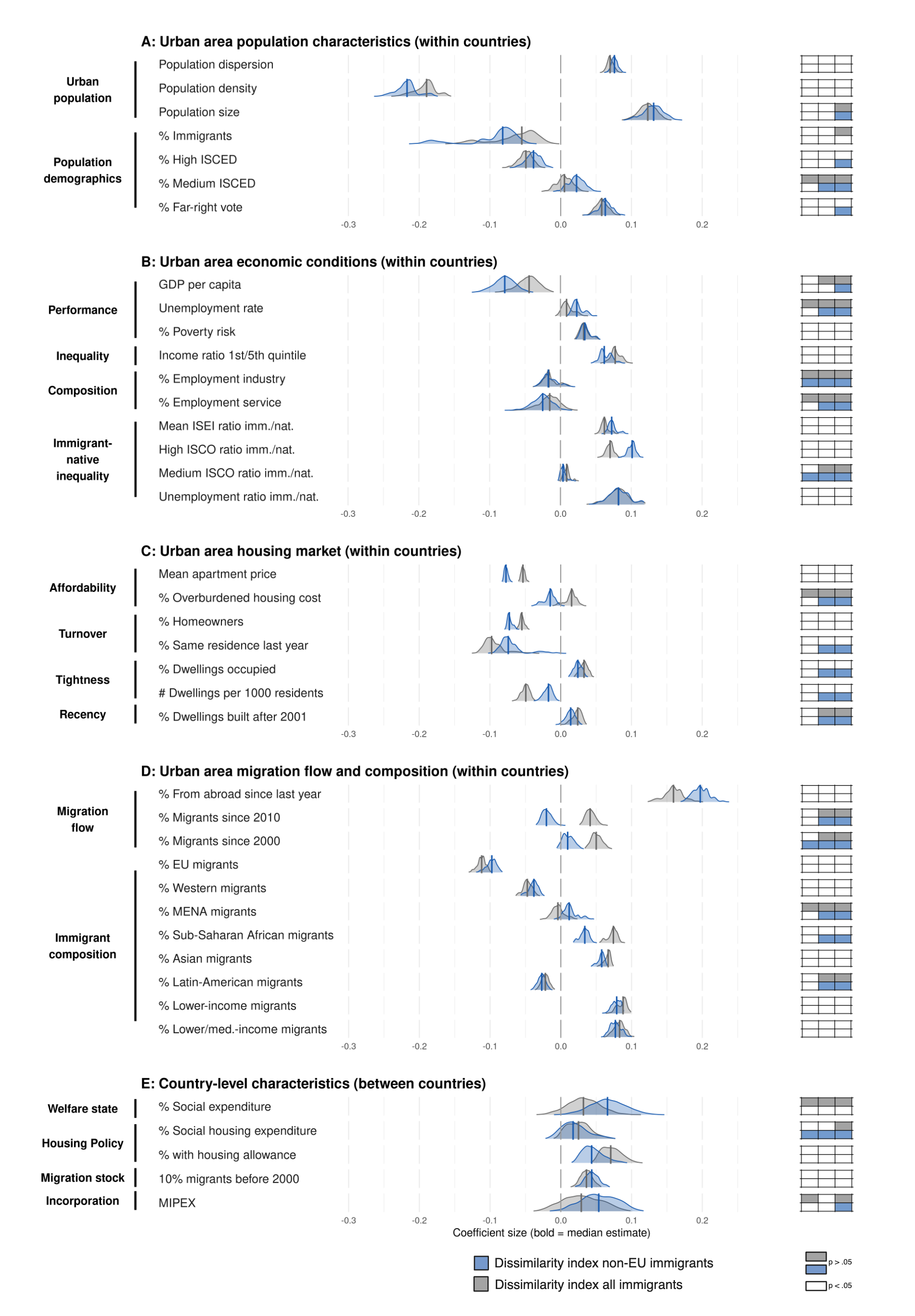}
  \caption{Results for dissimilarity index with non-EU immigrants only (blue) vs. all immigrants (grey).}
  \label{fig:c3_oth}
\end{figure}

\begin{figure}[h!]
  \centering
  \includegraphics[width=0.95\linewidth]{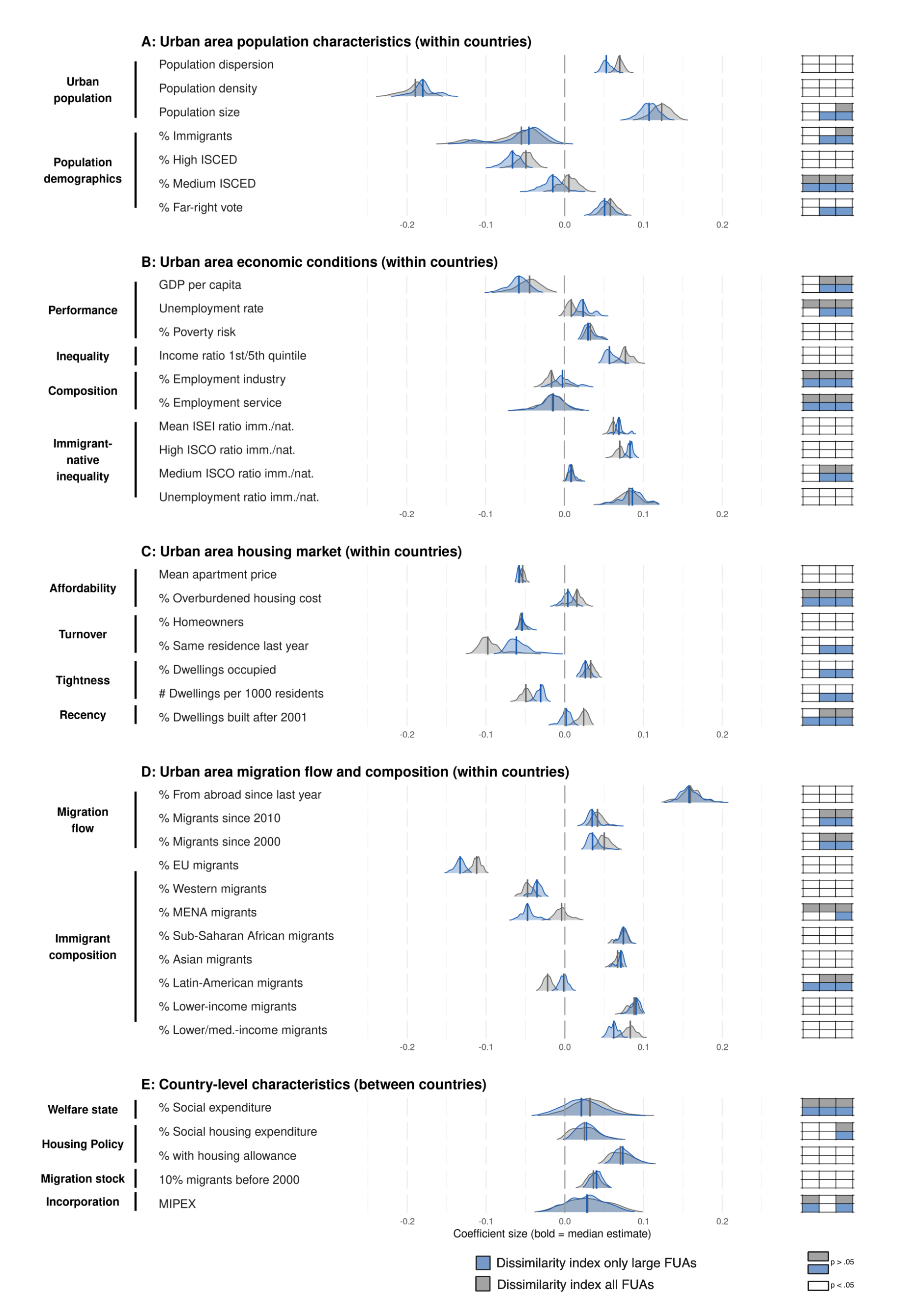}
  \caption{Results for dissimilarity index when excluding small FUAs with population < 100,000 (blue) vs. all FUAs (grey).}
  \label{fig:c4_large}
\end{figure}

\begin{figure}[h!]
  \centering
  \includegraphics[width=0.95\linewidth]{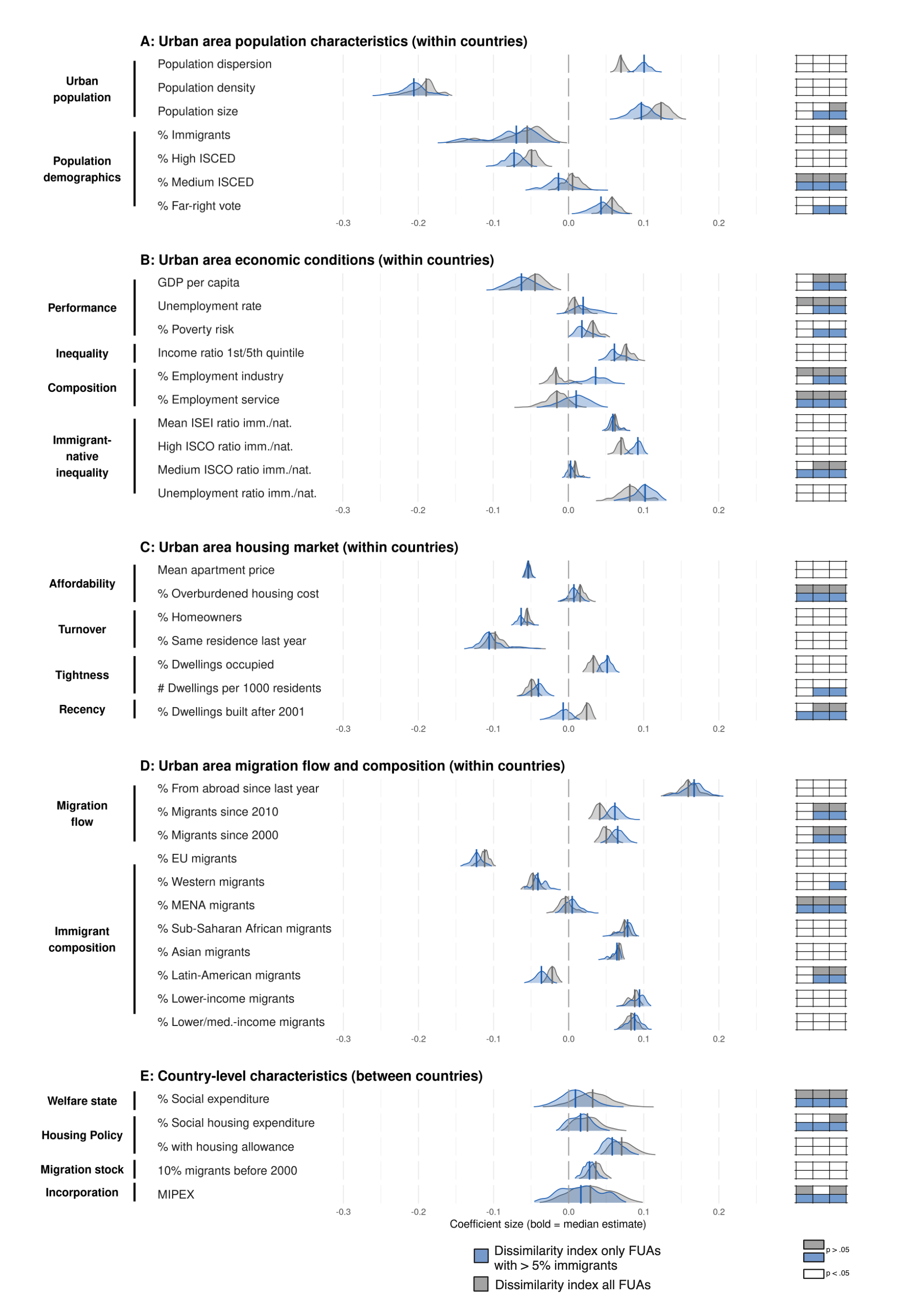}
  \caption{Results for dissimilarity index when excluding FUAs with proportion of immigrants < 5\% (blue) vs. all FUAs (grey).}
  \label{fig:c5_foreign}
\end{figure}

\begin{figure}[h!]
  \centering
  \includegraphics[width=0.95\linewidth]{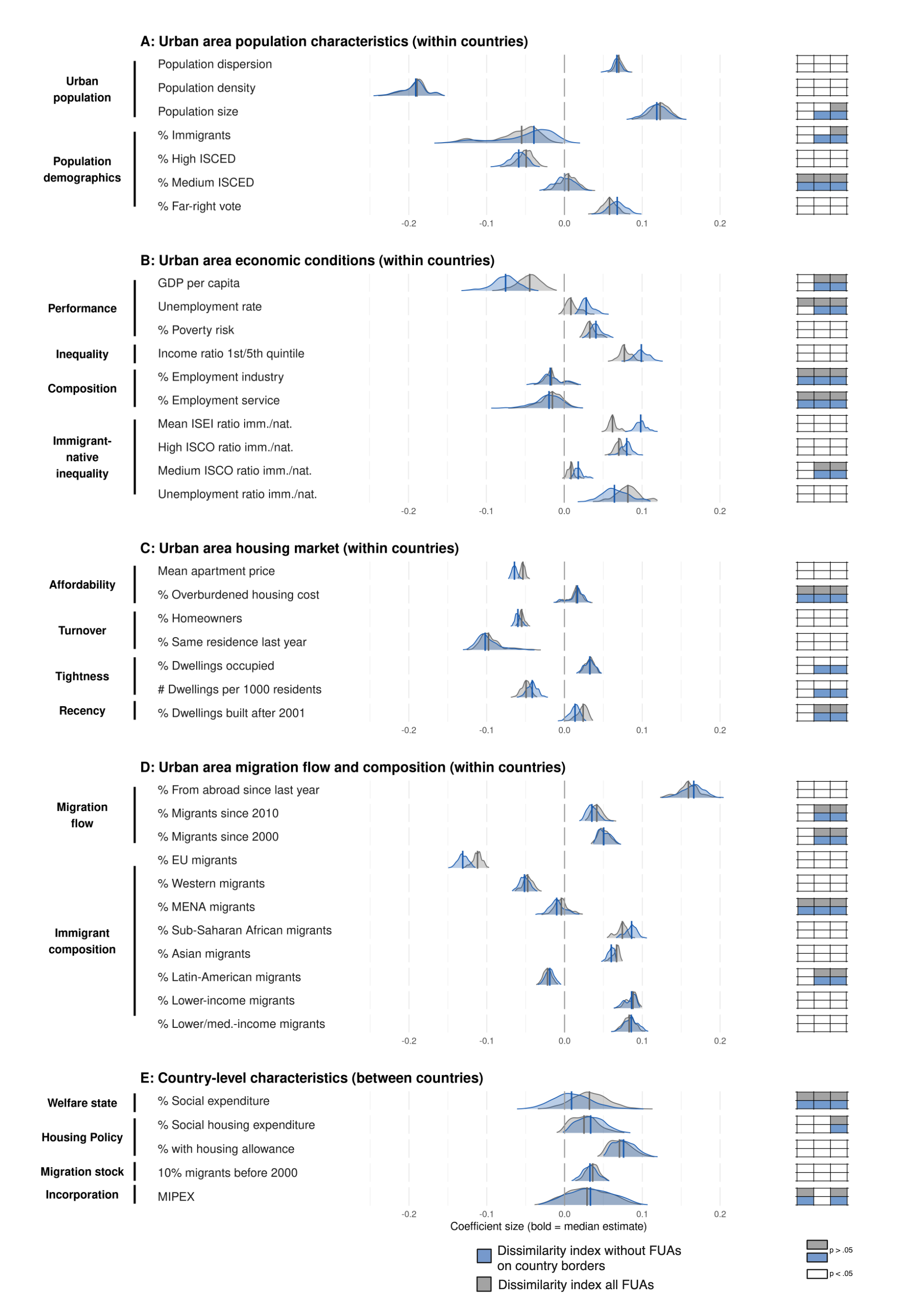}
  \caption{Results for dissimilarity index when excluding FUAs on country borders (blue) vs. all FUAs (grey).}
  \label{fig:c6_borders}
\end{figure}

\clearpage

\section{Robustness – immigrants and descendants} \label{suppl:D}

\begin{figure}[h!]
  \centering
  \includegraphics[width=1\linewidth]{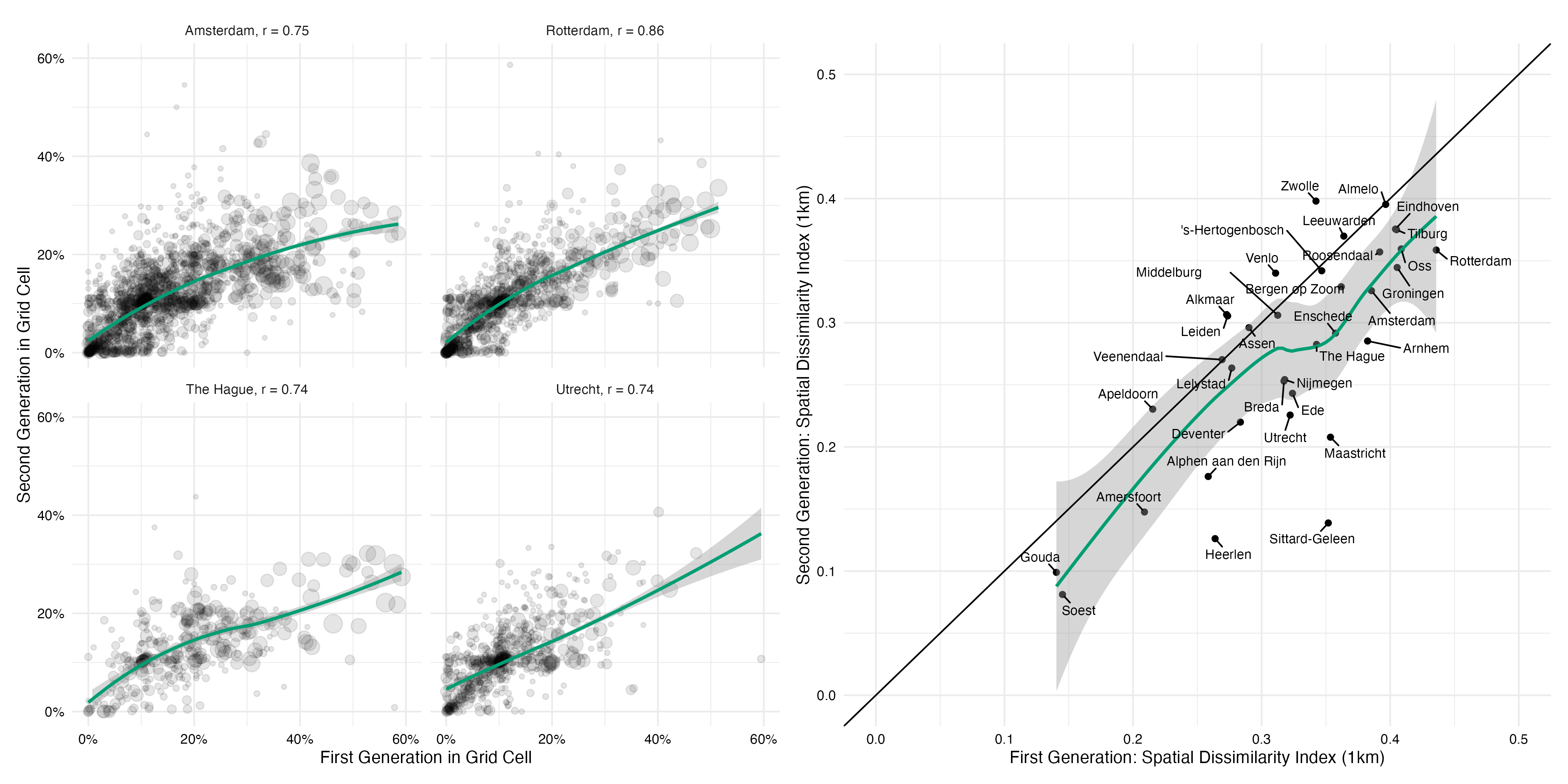}
  \caption{Segregation at 1km scale among the first and the second generation in the Netherlands. A) Scatterplots showing the proportion of the first- and second generation in each grid cell among the four largest FUAs of the Netherlands. B) Scatterplot showing the segregation index of the first- and the second generation from natives for all FUAs in the Netherlands.}
  \label{fig:d_nl}
\end{figure}

\clearpage

\section{Variable definitions} \label{suppl:E}

\begin{footnotesize}
\setlength{\tabcolsep}{6pt}
\renewcommand{\arraystretch}{1.1}
\begin{longtable}{p{1.8cm} p{1.8cm} p{2cm} p{5cm} p{1.4cm} p{1.4cm}}
\caption{Summary of variables used in the Specification Curve Analysis.}
\label{tab:variables}\\
\toprule
\textbf{Variable Group} & \textbf{Subgroup} & \textbf{Variable name} & \textbf{Variable description} & \textbf{Level} & \textbf{Year} \\
\midrule
\endfirsthead
\toprule
\textbf{Variable Group} & \textbf{Subgroup} & \textbf{Variable name} & \textbf{Variable description} & \textbf{Level} & \textbf{Year} \\
\midrule
\endhead
\midrule
\multicolumn{6}{r}{\textit{Continued on next page}}\\
\bottomrule
\endfoot
\bottomrule
\endlastfoot

\textbf{Population Characteristics (FUA)} & Urban Population & Population dispersion & Normalized Shannon entropy (population concentrated at low levels and dispersed at high levels). & FUA & 2021/22 \\
 &  & Population density & Residents per square km & FUA & 2021/22 \\
 &  & Population size & Number of residents & FUA & 2021/22 \\
 & Population demographics & \% Immigrants & Proportion of immigrants & FUA & 2021/22 \\
 &  & \% High education & Proportion 25--64 with highest education corresponding to ISCED levels 5--8 & FUA & 2018 \\
 &  & \% Medium education & Proportion 25--64 with highest education corresponding to ISCED levels 3--4 & FUA & 2018 \\
 &  & \% Far-right vote & \% of votes for parties belonging to the far-right block in 2019 European elections & NUTS3; 4 countries NUTS2 & 2019 \\

\textbf{Economic conditions (FUA)} & Performance & GDP per capita & GDP per capita & FUA & 2018 \\
 &  & Unemployment rate & Unemployment rate & FUA & 2018 \\
 &  & \% Poverty risk & Share of people with equivalised disposable income below 60\% of the national median & NUTS2 & 2018 \\
 & Inequality & Income ratio 1st/5th quintile & Ratio of total income received by 20\% of population with highest income to 20\% with the lowest income. & NUTS2 & 2018 \\
 & Composition & \% Employment industry & Percentage of population employed in industry/construction & NUTS3 & 2018 \\
 &  & \% Employment service & 1 minus percentage of population employed in industry, construction, agriculture, forestry, or fishing & NUTS3 & 2018 \\
 & native-immigrant inequality & Mean ISEI ratio imm./nat. & Ratio of mean ISEI in immigrant to native population. Mean ISEI calculated from average ISEI scores for ISCO-1 codes (Ganzeboom 2010) and proportion employed in ISCO-1-code occupations & NUTS2 & 2021/22 \\
 &  & High ISCO ratio imm./nat. & Ratio of proportion with high ISCO-1 codes in immigrant to native population. High ISCO codes encompass managers (ISCO 1000) and professionals (ISCO 2000) & NUTS2 & 2021/22 \\
 &  & Medium ISCO ratio imm./nat. & Ratio of proportion with medium ISCO-1 in immigrant to native population. Medium ISCO codes encompass technicians/associate professionals (ISCO 3000) and clerical support workers (ISCO 4000) & NUTS2 & 2021/22 \\
 &  & Unemployment ratio imm./nat. & Ratio of proportion of unemployed in the active immigrant population to proportion of unemployed in the active native population & NUTS2 & 2018 \\

\textbf{Housing Market (FUA)} & Affordability & Mean apartment price & Average price for buying an apartment & FUA & 2018 \\
 &  & \% overbudedend housing cost & Share of population in households with total housing costs exceeding 40\% of disposable income & NUTS2 & 2018 \\
 & Turnover & \% homeowners & Proportions of households in owned dwelling & NUTS3 & 2021/22 \\
 &  & \% same residence last year & Proportion of residents who lived in the current dwelling one year prior & FUA & 2021/22 \\
 & Tightness & \% dwellings occupied & Proportion of conventional dwellings that are occupied & NUTS3 & 2021/22 \\
 &  & Number dwellings per 1000 residents & Number of dwellings per 1000 residents & NUTS3 & 2021/22 \\
 & Recency & \% dwellings built after 2001 & Proportion of conventional dwellings built after 2001 & NUTS3 & 2021/22 \\

\textbf{Migrants' characteristics (FUA)} & Migration flow & \% from abroad since last year & Proportion of FUA population that moved to FUA from abroad since last year & FUA & 2021/22 \\
 &  & \% migrants since 2010 & Proportion who arrived after 2010 among all immigrants & NUTS3 & 2021/22 \\
 &  & \% migrants since 2000 & Proportion who arrived after 2000 among all immigrants & NUTS3 & 2021/22 \\
 & Immigrant composition & \% EU migrants & Proportion EU migrants among all migrants & FUA & 2021/22 \\
 &  & \% Western migrants &  & NUTS3 & 2021/22 \\
 &  & \% MENA migrants &  & NUTS3 & 2021/22 \\
 &  & \% Sub-saharan African migrants &  & NUTS3 & 2021/22 \\
 &  & \% Asian migrants &  & NUTS3 & 2021/22 \\
 &  & \% Latin-American migrants &  & NUTS3 & 2021/22 \\
 &  & \% Lower-income migrants &  & NUTS3 & 2021/22 \\
 &  & \% Lower-medium income migrants &  & NUTS3 & 2021/22 \\

\textbf{Country characteristics} & Welfare state & \% social expediture & Government expenditure on social welfare as \% of GDP & country & 2018 \\
 & Housing policy & \% social housing expenditure & Government expenditure on social housing as \% of GDP & country & 2018 \\
 &  & \% with housing allowance & Percentage of population receiving housing allowances & country & 2024 \\
 & Migration stock & 10\% immigrants before 2000 & 10\% immigrants among population before 2000 & country & 1990--2020 \\
 & Incorporation & MIPEX & Higher values indicate more rights for immigrants (in labour market, family reunion, education, health, political participation, permanent residence, access to nationality, anti-discrimination) & country & 2018 \\
\end{longtable}
\end{footnotesize}

\end{document}